\documentclass[fleqn,usenatbib]{mnras}
\pdfoutput=1
\usepackage[T1]{fontenc}
\usepackage{ae,aecompl}

\usepackage[english]{babel}
\usepackage{graphicx}
\usepackage{amsmath,mathtools}
\usepackage{amssymb}
\usepackage{bm}
\usepackage{xcolor}
\usepackage{pdflscape}
\usepackage{makecell}
\usepackage{multirow}
\usepackage{rotating}
\usepackage{orcidlink}

\usepackage{aas_macros}
\graphicspath{{fig}}

\renewcommand{\vec}[1]{{\bm{\mathrm{#1}}}}
\newcommand{\mathdash}{\,\text{---}\,}
\newcommand{\diff}[2]{{\frac{d{#1}}{d{#2}}}}
\newcommand{\pdiff}[2]{{\frac{\partial{#1}}{\partial{#2}}}}

\title{The realm of Aurora. Density distribution of metal-poor giants in the heart of the Galaxy}

\author[Kurbatov et al.]{
Evgeny P. Kurbatov$^1$\orcidlink{0000-0002-1024-9446}\thanks{E-mail: evgeny.kurbatov@ast.cam.ac.uk},
Vasily Belokurov$^1$\orcidlink{0000-0002-0038-9584}\thanks{E-mail: vasily@ast.cam.ac.uk},
Sergey Koposov$^{2,3}$\orcidlink{0000-0003-2644-135X},
Andrey Kravtsov$^4$\orcidlink{0000-0003-4307-634X},
\and
Elliot Y. Davies$^1$\orcidlink{0000-0001-5996-4072},
Anthony~G.~A. Brown$^5$,
Tristan Cantat-Gaudin$^6$,
Alfred Castro-Ginard$^5$,
\and
Andrew~R. Casey$^{7,8,9}$\orcidlink{0000-0003-0174-0564},
Ronald Drimmel$^{10}$,
Morgan Fouesneau$^6$,
Shourya Khanna$^{10}$,
\and
Hans-Walter Rix$^5$,
Alex Wallace,$^7$
\\
\\$^1$Institute of Astronomy, Madingley Rd, Cambridge, CB3 0HA, UK
\\$^2$Institute for Astronomy, University of Edinburgh, Royal Observatory, Blackford Hill, Edinburgh EH9 3HJ, UK
\\$^3$Kavli Institute for Cosmology, University of Cambridge, Madingley Road, Cambridge CB3 0HA, UK
\\$^4$Department of Astronomy and Astrophysics, The University of Chicago, Chicago, IL 60637, USA
\\$^5$Leiden Observatory, Leiden University, Einsteinweg 55, 2333 CC Leiden, The Netherlands
\\$^6$Max Planck Institute for Astronomy, K\"{ o}nigstuhl 17, 69117 Heidelberg, Germany
\\$^7$School of Physics and Astronomy, Monash University, Clayton VIC 3800, Australia
\\$^8$Center for Computational Astrophysics, Flatiron Institute, 162 5th Avenue, New York City 10010, New York, USA
\\$^9${Centre of Excellence for Astrophysics in Three Dimensions (ASTRO-3D)}
\\$^{10}$INAF - Osservatorio Astrofisico di Torino, via Osservatorio 20, 10025 Pino Torinese (TO), Italy
}

\begin{document}
\maketitle

\begin{abstract}
The innermost portions of the Milky Way's stellar halo have avoided scrutiny until recently. The lack of wide-area survey data, made it difficult to reconstruct an uninterrupted view of the density distribution of the metal-poor stars inside the Solar radius. In this study, we utilize red giant branch (RGB) stars from \textit{Gaia}, with metallicities estimated using spectrophotometry from \textit{Gaia} Data Release 3. Accounting for \textit{Gaia}'s selection function, we examine the spatial distribution of metal-poor ([M/H]$<-1.3$) RGB stars, from the Galactic center ($r \approx 1$ kpc) out to beyond the Solar radius ($r \approx 18$ kpc). Our best-fitting single-component cored power-law model shows a vertical flattening of $\approx 0.5$ and a slope $\approx -3.4$, consistent with previous studies. Motivated by the mounting evidence for two distinct stellar populations in the inner halo, we additionally test a range of two-component models. One of the components models the tidal debris from the \textit{Gaia} Sausage/Enceladus merger, while the other captures the Aurora population -- stars that predate the Galactic disk formation. Our best-fit two-component model suggests that both populations contribute equally around the Solar radius, but Aurora dominates the inner halo with a steeper power-law index of $\approx -4.5$, in agreement with the nitrogen-rich star distribution measured by \citet{Horta2021MNRAS.500.5462H}.
\end{abstract}

\begin{keywords}
Galaxy: centre -- Galaxy: structure -- Galaxy: evolution -- Galaxy: abundances -- methods: statistical
\end{keywords}

\section{Introduction}
\label{sec:intro}

Our Galaxy was small when it was young. Later, as the Galactic mass increased, the metal-poor stars born at high redshift, during the early phase of the Milky Way (MW) formation were pushed further in. By redshift zero, they have been engulfed by the prolific stellar mass build-up, concealed by the secular morphological transformations such as the disc emergence and the bar buckling, and enshrouded by the dust left over from the latest bouts of star formation. As a result, the stellar population tracing the primordial state of the MW has been hidden from our sight up until now. Most recently, the veil on the innermost prehistoric parts of the Galaxy has started to lift thanks to the data from the ESA \textit{Gaia} mission \citep[][]{GaiaCollaboration2016A&A...595A...1G} and high-resolution large spectroscopic surveys such as APOGEE \citep[][]{Majewski2017}.

For example, Pristine Inner Galaxy Survey \citep[PIGS][]{Arentsen2020} combined Gaia, Pan-STARRS and narrow-band photometry to identify candidate metal-poor red-giant stars in the direction of the Galactic bulge. They utilized Gaia parallaxes to restrict their sample to stars located near the Galactic center. PIGS stars were followed-up spectroscopically to measure metallicity [Fe/H] and line-of-sight velocity. \citet{Arentsen2020, Arentsen2020_2} see a clear evolution of the stellar kinematics in the inner MW: as the metallicity of the PIGS stars decreases, the rotational signal disappears and the velocity dispersion increases. The PIGS study culminated with the orbital analysis of the sample \citep[][]{Ardern-Arentsen2024} which demonstrated that most of the metal-poor stars studied remain confined to the inner $5$ kpc of the Galaxy's center. They also show evidence for two distinct metal-poor components among the PIGS stars in the range $-2<$[Fe/H]$<-1.5$: a more concentrated one with a slightly higher rotation velocity of $v_{\phi}\approx80$ km/s and a more diffuse one with a lower $v_{\phi}\approx40$ km/s. The faster component dominates also for [Fe/H]$>-1.5$ and the slower one dominates for [Fe/H]$<-2$. These PIGS results are consistent with the measurements obtained by the Chemical Origins of Metal-poor Bulge Stars survey (COMBS) who selected their targets from narrow-band SkyMapper photometry \citep[][]{Lucey2019,Lucey2021}.

Aided by its infrared vision, APOGEE can peer through the dust and thus has been used extensively to study stellar populations in the inner Galaxy. \citet{Schiavon2017} relied on APOGEE's high quality abundance measurements to discover a population of stars with enhanced [N/Fe] and [Al/Fe] characteristic of Globular Clusters, located a few kpc from the center of the Milky Way. \citet{Horta2021MNRAS.500.5462H} focused on the metal-poor portion of the N-rich population and taking the APOGEE's selection function into account, inferred their spatial density distribution. \citet{Horta2021MNRAS.500.5462H} find that these stars, presumably with Globular Cluster (GC) origin, follow a very steep radial density profile, with a power-law index $\alpha\approx-4.5$ and a vertical flattening $q\approx0.5$.

In the observational analyses above, the central Galactic region studied is referred to as "bulge", partly for historical reasons, partly to acknowledge its non-disc appearance. While often not stated explicitly, it is assumed that the "bulge" region as a whole contains a mixture of structural components. At high metallicities it is principally made up by the bar and the disc \citep[but see the recent mentions of the small and metal-rich "knot" in the MW centre][]{Horta2024, Rix2024}, and at low [Fe/H] by the stellar halo with a quasi-spheroidal shape, which, as evidenced from the stellar kinematics, is dispersion-supported. Ironically, in the Galactic "bulge" region, the existence of an actual, classical bulge, i.e. an old but metal-rich spheroid with low angular momentum is contested. Furthermore, the exact genesis of the inner stellar halo has also remained unclear until lately. However, \citet{Belokurov2022MNRAS.514..689B} have recently proposed that the inner halo may contain a substantial, previously overlooked stellar population called {\it Aurora}, which formed within the MW itself at high redshift, prior to the Galactic disc emergence. By comparing to numerical simulations of galaxy formation, \citet{Belokurov2022MNRAS.514..689B} conclude that although first detected in the extended Solar neighborhood \citep[see also][for an independent study using H3 survey]{Conroy2022}, Aurora ought to follow a steep Galactocentric radial density profile, with most of its stars in the central few kpc of the MW.

\begin{figure*}
  \centering
  \includegraphics[width=\textwidth]{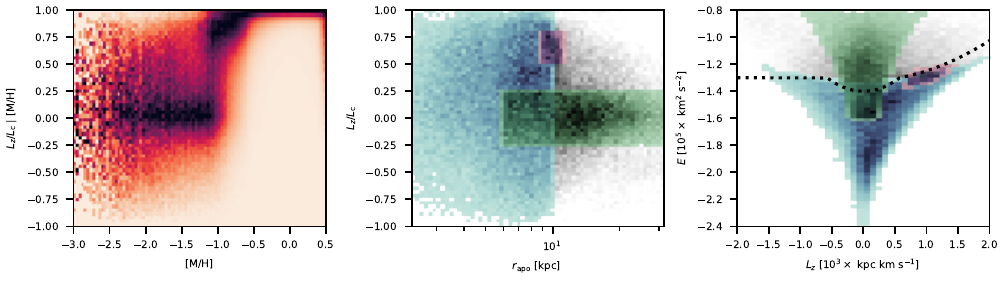}
  \caption{Orbital properties of the vetted RGB sample in \citet{Andrae2023ApJS..267....8A}. {\bf Left:} Column-normalized distribution of orbital circularity vs. metallicity. {\bf Middle:} orbital circularity vs. apocentric distance. {\bf Right:} energy vs. the vertical component of the angular momentum. In the last two panels, stars are subjected to a metallicity selection $-2.0 < [\mathrm{M}/\mathrm{H}] \leq -1.5$. The approximate locations of two halo populations are marked in colour: the Aurora stars are in blue, and the GS/E stars are in green. The stars likely captured by bar resonances  are shown in magenta. See Section \ref{sec:sample-selection} for further details. The dotted black line in the right panel shows the boundary  separating the in-situ and accreted populations according to \citet{Belokurov2023}.}
  \label{fig:data-kinem-all}
\end{figure*}

Aurora's chemical make-up is distinct from the rest of the in-situ stars as revealed by     its elevated elemental abundance spread \citep{Belokurov2022MNRAS.514..689B}. This abundance variance is most pronounced for elements such as Mg, Al, N, Si, O, i.e. those that are also known to show anomalously large scatter in the surviving MW globular clusters \citep[see e.g.][]{Gratton2004, Bastian2018, Gratton2019}. Using [N/O] as a chemical fingerprint of the GC origin, \citet{Belokurov2023} show that the orbital energy distributions of the Aurora stars and stars from disrupted GCs are indistinguishable. \citet{Belokurov2023} track the formation of the MW in-situ GCs as a function of metallicity and show that the GC contribution to the Galactic star formation (SF) was at its highest in the pre-disc, Aurora era when it could reach values as high as $50\%$. Such high SF share in bound massive star clusters is highly unusual for today's Galaxy but may well have been typical for galaxies at high redshifts \citep[see e.g.][]{Mowla2024}. \citet{Belokurov2023} conclude that the radial density profiles of the Aurora and the high-[N/O] field stars are likely very similar, i.e. a steep power-law with an index of order of $-4$ as estimated by \citet{Horta2021MNRAS.500.5462H}. Assuming the GC contribution to SF and the initial GC masses computed in \citet{Baumgardt2003}, \citet{Belokurov2023} give an estimate of the total stellar mass of Aurora of $\sim5\times10^8M_{\odot}$.

Several other non-disc Galactic components are expected to inhabit the inner Milky Way. One of these is the so-called Gaia-Sausage/Enceladus (GS/E), the tidal debris cloud left behind by the last major merger, an encounter between the MW and a massive dwarf galaxy some 10 Gyr ago \citep[][]{Belokurov2018MNRAS.478..611B, Helmi2018}. First glimpses of the GS/E's contribution to the Galactic halo density profile were identified by \citet{Deason2013} who interpreted the break in the halo's radial density profile \citep[e.g.][]{Watkins2009, Deason2011, Sesar2011} as the evidence for the apo-centric pile-up of tidal debris deposited by a single, massive ancient event \citep[also see][for the follow-up]{Deason2018pileup}. These earlier studies reported a shallower power-law with an index $\sim -2.6$ in the inner Galaxy, within Galactocentric distances $20$-$30$ kpc, and a steeper one with an index $\sim -4.5$ further out. Note also that several studies argued for a change in the halo's flattening with radius as opposed to a change in the radial density steepness \citep[see e.g.][]{Xue2015, Iorio2018, Iorioa2019}.

Note that the earlier studies described above were concerned with the overall halo density distribution rather than the behaviour of the GS/E debris. While the GS/E stars may dominate the halo around the Solar neighborhood \citep[][]{Belokurov2018MNRAS.478..611B}, its exact contribution in the inner MW had remained unconstrained until recently. The heart of the Galaxy, i.e. the region at low Galactocentric radii $r<5$ kpc and/or low heights $|z|<5$ kpc had always been difficult to observe and thus it had usually been excluded from the modelling efforts. Most recently, attempts have been made to leverage the power of the APOGEE infra-red data to decipher the GS/E's density distribution across a wide range of radii, including close to the very centre of the Galaxy \citep[][]{Mackereth2020MNRAS.492.3631M, Lane2023MNRAS.526.1209L}. In particular, \citet{Lane2023MNRAS.526.1209L} show that the radial density distribution of the GS/E's stars might be doubly broken, revealing an even shallower density in the inner Galaxy. They find a power-law index close to $-1$ inside $\sim 10$ kpc, $-3$ between $10$ and $30$ kpc and $-6$ further out. These results are broadly consistent with the study of \citet{Han2022} who use spectroscopic data from the H3 survey \citep[][]{Conroy2019} to find the power-law index changing from $-1.7$ to $-3.1$, and then to $-4.6$ at similar radii.

While there is no strong evidence that the Galaxy experienced significant mergers after colliding with the GS/E's progenitor \citep[but see][for an alternative view]{Donlon2022}, the accretion history before the GS/E's arrival is more unconstrained. Based on the APOGEE data, \citet{Horta2021heracles} argue that the inner MW is swamped with the tidal debris from an ancient merger event, predating that of the GS/E. They estimate the stellar mass of the progenitor galaxy (dubbed Heracles) to be of order of $5\times10^8 M_{\odot}$ and show that Heracles stars are chemically distinct from the GS/E stars. The Heracles conjecture resonates with an earlier hypothesis of the existence of the so-called Kraken galaxy (also known as Koala), revealed by a group of GCs with low orbital energies \citep[][]{Massari2019, Kruijssen2020, Forbes2020}. Revisiting the accreted/in-situ classification of the Galactic GCs, \citet{Belokurov2024gc} assign most of the Kraken/Koala GCs to the MW Aurora population, but point out that some $\sim10\%$ of the low-energy GCs can be contributed by accretion events.
Originally, the spatial extent of the Heracles debris was estimated to be limited to within $\sim 4$ kpc of the Galactic centre. However, taking the effects of the APOGEE selection function into account \citep[see][]{Lane2022}, it is likely the spatial extent of this population is larger than previously thought. At a first glance, the Aurora and the Heracles populations appear very similar to each other. Although currently it is difficult to compare faithfully their chemical fingerprints given mutually exclusive selection cuts applied, there is plenty of evidence that there is a very substantial overlap between the two in the abundance space \citep[][]{Myeong2022,Horta2023}.

\begin{figure*}
  \centering
  \includegraphics[width=0.95\textwidth]{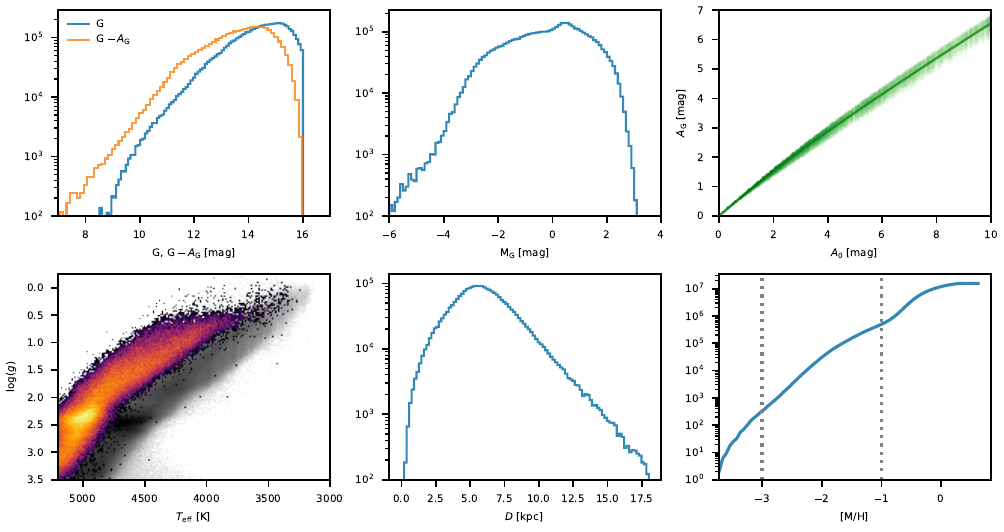}
  \caption{Photometric, spatial and chemical properties of the RGB sample. {\bf Top left:} Distribution function of apparent magnitude $G$ of the low-metallicity ($-3 < [\mathrm{M}/\mathrm{H}] < -1$) subsample of RGB stars. \textit{Top center:} Same as the previous panel but for the absolute magnitude $M_G$. {\bf Top right:} Polynomial fit of the $\mathrm{G}$ band extinction as a function of the median monochromatic extinction $A_0$. Green shaded area shows all the stars in the catalogue of RGB stars \citep{Andrae2023ApJS..267....8A}. {\bf Bottom left:} Kiel diagram (effective temperature vs surface gravity) of the low-metallicity RGB stars (red-and-yellow density map) and all RGB stars (grey density map in the background). {\bf Bottom center:} Distribution for the Heliocentric distance. {\bf Bottom right:} Cumulative distribution of the stellar metallicity in the RGB sample. The vertical dotted lines are the metallicity limits used in our analysis.}
  \label{fig:data-magnitudes}
\end{figure*}

RR Lyrae are a unique halo tracer able to penetrate the dusty mess of the inner MW, thanks to their easily recognisable pulsation pattern and a relatively high intrinsic luminosity. \citet{Pietru2015} measure a vertical flattening of $\approx0.5$ and a power-law index of $\approx-3$, as deprojected by \citet{Perez2017},  in the inner $\sim 1$ kpc of the Galaxy with RR Lyrae identified in the OGLE survey \citep[][]{Soszynski2014}. \citet{Perez2017} show that when extended outwards, the inner RR Lyrae density profile $\propto r^{-3}$ agrees reasonably well (in both the slope and the normalisation) with the halo radial density measurements between $5$ and $2$ kpc. These measurements are also in agreement with the study of \citet{Wegg2019} who rely on the RR Lyrae detected with PanSTARRs \citep[][]{Sesar2017} to show that between $2$ and $20$ kpc from the Galactic centre, the halo's density follows a power-law with an index of $-2.6$.

\begin{figure*}
  \centering
  \includegraphics[width=\textwidth]{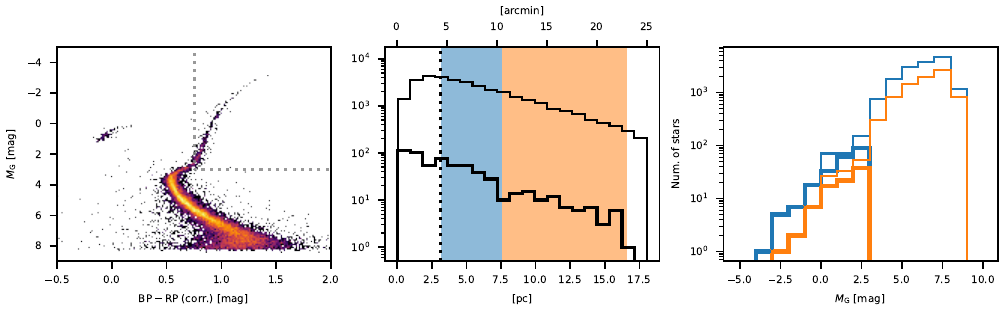}
  \caption{Estimate of the stellar luminosity function at low metallicity. {\bf Left:} Extinction-corrected colour-magnitude diagram of the globular cluster NGC~6397. Grey dotted lines mark the boundaries of the RGB selection. {\bf Middle:} Histograms of the cluster-centric radial distributions of RGB stars (thick lines) and all cluster stars (thin lines). Dotted vertical line is the half-mass radius of the cluster. Blue and orange shaded area are radial distance ranges used to estimate stellar luminosity functions. {\bf Right:} Luminosity Functions of the RGB stars (thick lines) and all the stars (thin lines) for the inner (blue) and outer (orange) radial layers.}
  \label{fig:ngc-6397}
\end{figure*}

The RR Lyrae density (power law index between $-3$ and $-2.6$) is steeper than the radial profile of the GS/E debris (power law index $>-2$) but shallower than the inferred density of the GC-born (N-rich) stars (power law index $<-4$). At face value, this seems fine if the reported RR Lyrae density is an average of these distinct populations. While to date no attempt has been made to measure the exact contributions of individual components to the Galactic halo density with RR Lyrae, several studies showed that such dissimilar populations are indeed traceable with RR Lyrae. For example, \citet{Belokurov2018unmixing} show that the fractional contribution of the so-called Oosterhoff Type I and II \citep[][]{Oosterhoff1939,Oosterhoff1944, Catelan2004, Catelan2009} and the High Amplitude Short Period \citep[HASP,][]{Stetson2014,Fiorentino2015} RR Lyrae changes dramatically as a function of Galactocentric radius. The ratio of type I/II evolves on the same spatial scale as the halo radial density break and is likely associated with the evolving contribution of the GS/E debris to the halo overall.

The HASP fraction on the other hand shows two distinct components, one with the radial scale of $20$-$30$ kpc is probably related to the GS/E, However, the largest fraction of HASP RR Lyrae is detected in the inner $5$-$10$ kpc of the MW as a compact and dense population unrelated to the GS/E. \citet{Iorio2021} show that the Galactic RR Lyrae, as identified and measured by \textit{Gaia}, can be split into three distinct groups based on their kinematics. In the radial range $5<r$(kpc)$<25$, between $50\%$ and $80\%$ can be attributed to the GS/E given their high radial anisotropy, with orbital anisotropy $\beta\approx0.9$. The remaining RR Lyrae are either in a rapidly rotating and highly flattened disc population or in a quasi-isotropic, non-rotating halo component. The isotropic component contains a dense, centrally concentrated "core" with an elevated HASP fraction. Note that high HASP fractions are linked to the most massive progenitor galaxies, moreover, the HASP ratio in the inner MW exceeds that observed in the largest Galactic satellites, the LMC and the SMC \citep[][]{Fiorentino2015, Belokurov2018unmixing}. Therefore, it is quite likely that the dense inner population of RR Lyrae with quasi-isotropic kinematics and elevated HASP fraction is related to the in-situ, pre-disc Aurora component.

Most recently, thanks to the \textit{Gaia} Data Release 3 \citep[][]{GaiaCollaboration2023A&A...674A...1G}, it has become possible to build panoramic maps of the Galactic metal-poor population. Using {\it Gaia's} XP spectro-photometry \citep[][]{Montegriffo2023, DeAngeli2023A&A...674A...2D}, several groups endeavoured to estimate stellar metallicities with reasonable success \cite[][]{Rix2022, Andrae2023ApJS..267....8A, Bellazzini2023, Zhang2023MNRAS.524.1855Z, Yao2024, Hattori2024, Xylakis2024}. \citet{Rix2022} use their \textit{Gaia} XP-based metallicities to show that the distribution of metal-poor stars, i.e. those with [M/H]$<-1.5$ is centrally concentrated, tracing out the "poor old heart" (POH) of the Milky Way. \citet{Rix2022} estimate the total stellar mass in the POH component to be of order of $10^8 M_{\odot}$ and argue based on the age-metallicity relation presented in \citet{Xiang2022} that this Galactic component must be older than 12.5 Gyr. Note that this age estimate of the Galactic disc emergence agrees well with the calculation presented in \citet{Belokurov2024gc} based on Galactic GCs. \citet{Rix2022} point out that the POH stars must represent the tightly bound, low-energy old and (relatively) metal-poor spheroid whose outskirts were glimpsed closer to the Sun in the studies of \citet{Belokurov2022MNRAS.514..689B} and \citet{Conroy2022} and whose kinematics were scrutinised through the pencil beams of the PIGS and COMBS surveys \citep[][]{Ardern-Arentsen2024, Lucey2022}.

In this paper, we model the spatial distribution of Red Giant Branch stars with available \textit{Gaia} XP data to reveal the density distribution of the metal-poor population of the Galaxy. The manuscript is structured as follows. Section \ref{sec:sample-selection} discusses the dataset and selection criteria. Section \ref{sec:statistical-model} presents a general statistical model for the density distribution of the modeled population. One- and two-component models are described in Sections \ref{sec:spl-nogse} and \ref{sec:dpl}, respectively. The results are discussed in Section \ref{sec:discussion}. Details of the model calculations are provided in Appendices \ref{sec:transformation}, \ref{sec:parallax-error}, and \ref{sec:xp-selection}. Appendix \ref{sec:add-figs} contains additional figures.

\section{Sample selection and preparation}
\label{sec:sample-selection}

Our work is based on the vetted sample of RGB stars of \cite{Andrae2023Zenodo7945154} as described in \cite{Andrae2023ApJS..267....8A}. The sample consists of more than $17.5\,$M stars from the Gaia DR3 catalogue \citep{GaiaCollaboration2016A&A...595A...1G, GaiaCollaboration2023A&A...674A...1G} with measured XP spectra \citep{DeAngeli2023A&A...674A...2D}. All stars in this catalogue have effective temperature, surface gravity, and metallicity values estimated using the XGBoost model. The vetted RGB stars sample was created by applying a number of selection conditions designed to reduce contamination among the low-metallicity stars. These conditions affect the apparent magnitude ($\mathrm{G} < 16$), parallax and its error ($\varpi > 4\sigma_\varpi$), surface gravity ($\log_{10} g < 3.5$), effective temperature ($T_\mathrm{eff} < 5200\,$K), and also \textit{Gaia} and CatWISE colors \citep[see Sec.~4.2 of][]{Andrae2023ApJS..267....8A}.

\begin{figure*}
  \centering
  \includegraphics[width=\textwidth]{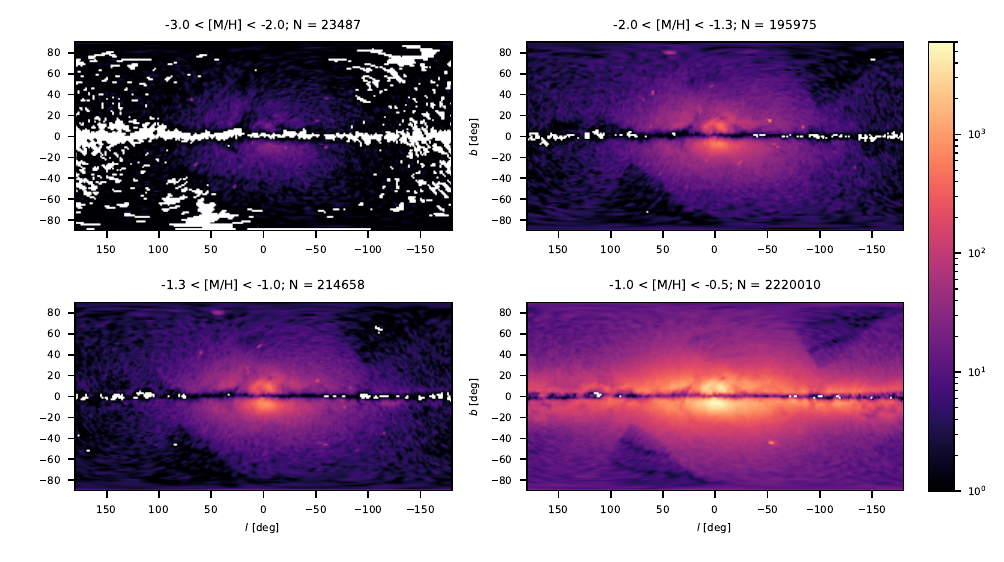}
  \caption{Counts of the RGB stars on the sky binned with HEALPix order 5. For each panel we give, $N$ the number of stars in the corresponding sample in the caption. Note the change in appearance of the projected density distribution from spheroidal at low metallicities to a disc-like at high metallicity (bottom right panel).}
  \label{fig:data-obs-sky}
\end{figure*}

Our primary sample consists of $433\,916$ stars with metallicities $-3.0 < [\mathrm{M}/\mathrm{H}] < -1.0$. This population is not uniform in its origin and kinematics. In order to distinguish between different components of the population under study, the following kinematic properties of the stars have been estimated: total energy $E$, azimuthal angular momentum $L_z$, the same for a perfect circular orbit $L_\mathrm{c}(E)$, and apocentric distance $r_\mathrm{apo}$. Transformation between Galactic and Galactocentric Cartesian coordinates of stars is performed assuming the Galactocentric position of the Sun in the Galaxy is $(-8.12, 0.0, 0.02)$ kpc \citep{GRAVITY2018A&A...615L..15G, Bennett2019MNRAS.482.1417B}. The Galactocentric solar velocities were set to $(12.9, 245.6, 7.78)$ km/s \citep{Drimmel2018RNAAS...2..210D, GRAVITY2018A&A...615L..15G, Reid2004ApJ...616..872R}. We used the \cite{Hunter2024arXiv240318000H} model for the Galactic potential, as implemented in the \textsc{agama} code \citep{Vasiliev2019MNRAS.482.1525V}\footnote{%
Specifically, the \texttt{MWPotentialHunter2024\_rotating} triaxial model with bar but no spiral arms.}. Figure~\ref{fig:data-kinem-all} presents chemo-dynamic properties of our sample. From left to right, the Figures shows i) the dependence of stellar circularity, i.e. the ratio of the vertical component of the angular momentum to that of the circular orbit $L_z/L_c$, on metallicity $[\mathrm{M}/\mathrm{H}]$ ii) circularity as a function of the apocentric distance, and finally, iii) total energy $E$ as a function of $L_z$.  As seen in the left panel of Figure \ref{fig:data-kinem-all}, the more metal-rich stellar population has a high circularity, consistent with a dominant, fast-rotating disc component. The rotation signature quickly disappears at lower metallicities in agreement with the earlier reported observations \citep{Belokurov2022MNRAS.514..689B,Zhang2023arXiv231109294Z, Chandra2023arXiv231013050C} as well as numerical models \citep{Dillamore2024MNRAS.527.7070D, Semenov2024ApJ...962...84S}. 

The overall kinematic and orbital behaviour of low-metallicity population is very different. If we limit the stars to metallicities $[\mathrm{M}/\mathrm{H}] \leq -1.5$, the rotation almost completely disappears and the dispersion in circularity shoots up. In this metallicity regime, two main components have been argued to dominate the volume close to the Sun: the {\it Gaia} Sausage/Enceladus (GS/E) merger debris, and the Aurora population \citep[][]{Myeong2022}. Note, however, as discussed earlier, that at very low  $[\mathrm{M}/\mathrm{H}]$, the dominance of both GS/E and Aurora likely declines giving way to the mix of debris from accreted low-mass sub-systems \citep[see discussion in e.g.][]{Belokurov2018MNRAS.478..611B, Zhang2023arXiv231109294Z}. Equally, the balance between different components close to the Galactic centre may be rather different from that near the Sun \citep[][]{Ardern-Arentsen2024}.

In the middle and right panels of Figure \ref{fig:data-kinem-all}) we separate crudely the stars with  $-2.0 < [\mathrm{M}/\mathrm{H}] \leq -1.5$ into high-eccentricity high-energy GS/E components (shaded green) and high-circularity-dispersion low-energy Aurora component (shaded blue). As the middle panel of the Figure illustartes, the GS/E's highly elongated orbits tend to reach large apocentric distances (up to $\sim 30$ kpc) while, naturally, the low-energy Aurora population stays closer to the Galactic centre. The selection into green and blue regions is done in the space of circularity and apocentric distance, but it roughly agrees with the $E, L_z$ boundary proposed by \citet{Belokurov2023,Belokurov2024gc} to separate the MW's accreted and in-situ populations (see right panel of the Figure). We also note a small prominent overdensity visible in Figure \ref{fig:data-kinem-all} and shaded purple. This sub-population has a strong preference for a particular energy or apocentric distance as well as the orbital circulatity. We surmise, echoing \citet{Dillamore2023MNRAS.524.3596D} that these stars have most likely been captured by the Galactic bar resonances. Please note that Figure \ref{fig:data-kinem-all} is used here for illustration purposes only as 
we are not planning to include dynamics of the Galaxy in our model presented below. 

To compute absolute magnitude values of the stars, we estimate the sky distribution of the monochromatic extinction at $541.4$~nm, $A_0$ \citep{Fitzpatrick1999PASP..111...63F, Delchambre2023A&A...674A..31D}, on a HEALPix 7 level using the \textsc{dustmaps} package \citep{Green2018JOSS}, and then apply a 3$^\text{rd}$-order polynomial approximation to get the extinction $A_\mathrm{G}$ in the Gaia $\mathrm{G}$ band as a function of $A_0$ and effective temperature $T_\mathrm{eff}$ \citep{Fitzpatrick2019ApJ...886..108F}. Parallaxes (in mas) are used as a distance measure (kpc). The absolute magnitudes in the $\mathrm{G}$ band are $\mathrm{M}_\mathrm{G} = \mathrm{G} - A_\mathrm{G} + 5\log_{10} \varpi - 10$. The observed (such as $\mathrm{G}$) and derived (such as $M_\mathrm{G}$ and $[\mathrm{M}/\mathrm{H}]$) properties of stars with $-3 < [\mathrm{M}/\mathrm{H}] \leq -1$ are shown in Figure \ref{fig:data-magnitudes}. As seen in the Figure, our selection (in $\mathrm{G}$, $\varpi$ and $T_\mathrm{eff}$ ) results in a very small number of stars brighter than $\mathrm{G} = 8$ and a lack of stars with $M_\mathrm{G} > 3$.

Because the observed luminosity function (LF) shown in the upper middle panel of Figure \ref{fig:data-magnitudes} carries  the effects of our selection, it can not represent the true luminosity distribution of stars in the central part of the Galaxy. To approximate the true LF, we use photometric data of a subsample of RGB stars in a representative globular cluster (GC), since the distances to clusters are well known usually. For our reference stellar population, we have chosen the NGC~6397 globular cluster, which is only $2.48$ kpc from the Sun (and hence its star counts are likely to be more complete)  and has a metallicity of $-2.0$ \citep{Vasiliev2021Zenodo4891252, Vasiliev2021MNRAS.505.5978V, Baumgardt2023MNRAS.521.3991B}. We estimate the luminosity function by counting stars in two GC-centric radial ranges (marked in blue and orange in the middle pane of Figure \ref{fig:ngc-6397}, both outside the half-mass radius of the cluster (Figure \ref{fig:ngc-6397}). These ranges are chosen to avoid the cluster's central parts probably suffering by blending. The middle and the right panels of the Figure show that reassuringly the LFs from the two cluster regions are consistent with each other. Therefore, for subsequent modelling, we use the LF of RGB stars from the inner radial layer (blue shaded area and the blue thick line on the Figure \ref{fig:ngc-6397}), because it contains a larger number of the bright stars. The LF has a power law-like shape and is well populated with the faint stars.

The sky distributions of the stars in four low-metallicity samples are shown on Figure \ref{fig:data-obs-sky}. As discussed in e.g. \citet{Rix2022}, the spatial distribution of stars evolves dramatically with metallicity. Even though the Figure focuses on the low-metallicity regime, the difference between the lowest-[M/H] (top left) and the highest-[M/H] (bottom right) sub-samples is clear. Stars with $-3 < [\mathrm{M}/\mathrm{H}] \leq -2$ are more centrally concentrated overall, whilst stars with $-1.3 < [\mathrm{M}/\mathrm{H}] \leq -1$ are more extended and flattened in the vertical direction. Note the bright spots at high latitudes. These are probably the globular clusters. They don't take up much space, so they shouldn't bias the distribution of RGB stars in the sample.

Unsurprisingly, our samples are incomplete near the Galactic plane. This depletion is due to the dust extinction \citep{Delchambre2023A&A...674A..31D}, but primarily it is caused by the minimum number of observations required for a star to be included in the \textit{Gaia} XP catalogue. This requirement also results in a lack of sources around the ecliptic band in areas that were scanned less frequently by Gaia \citep[Fig. 29]{Boubert2020MNRAS.497.1826B, DeAngeli2023A&A...674A...2D}. For example, see the 'triangles' in the lower-left and upper-right corners of the maps corresponding to declination angles $|\delta| \lesssim 20^\circ$. Below, we address both the dust extinction and scanning effects using a unified approach based on a sub-sample selection function.

\section{Statistical model for density distribution}
\label{sec:statistical-model}

\subsection{Density model}

Our goal is to recover the intrinsic spatial distribution of the Aurora stars using the low-metallicity RGB sample. We explore two distinct approaches to this task. First, we approximate the observed stellar distribution with a one-component model. Second, we fit a two-component model with a predefined contribution from the GS/E stars, based on the model of \cite{Lane2023MNRAS.526.1209L}.

In models commonly used for such problems, the density distributions are typically assumed to be axisymmetric and dependent on a dimensionless coordinate.
\begin{equation}
  m = \left( \frac{x^2 + y^2}{a^2} + \frac{z^2}{b^2} \right)^{\!1/2}  \;,
\end{equation}
where $x$, $y$ and $z$ are Galactocentric coordinates; $a$ and $b$ are semi-axes of a spheroid. Here we use two density profiles, namely i) a single power law with a sharp (but finite) core, ii) a broken power law with a flat core. The single power law is defined as follows:
\begin{equation}
  \label{eq:density-spl}
  \hat{\rho}_0 = \frac{1}{(1 + m)^{k_1}}  \;.
\end{equation}
Parameters of this model are the semi-axes $a$ and $b$, and the power index $k_1$. A special feature of the distribution \eqref{eq:density-spl} is that it is not flat at the centre but has a finite-amplitude peak. Indeed, at the coordinate origin the profile \eqref{eq:density-spl} behaves as $\hat{\rho}_0 \approx 1 - k_1 |x|/a$ in $x$-direction and as $\hat{\rho}_0 \approx 1 - k_1 |z|/b$ in $z$-direction. More generally, the approximate behavior of the profile \eqref{eq:density-spl} close to the origin is
\begin{equation}
  \hat{\rho}_0 \approx 1 - k_1 m  \;.
\end{equation}
Beyond the semi-axes scales, the distribution takes the form of the power law: $\hat{\rho}_0 \sim m^{-k_1}$.

The double power law is
\begin{equation}
  \label{eq:density-dpl}
  \hat{\rho}_0
  = \left\{
  \begin{array}{ll}
     \cfrac{1}{(1 + m^2)^{k_1/2}} & \text{if~} m \leq m_1  \;,  \\[8pt]
     \cfrac{m_1^{k_2} m^{-k_2}}{(1 + m_1^2)^{k_1/2}} & \text{if~} m > m_1  \;.
  \end{array}
  \right.
\end{equation}
Parameters of these model distributions are: semi-axes $a$ and $b$, the power indices $k_1$ and $k_2$ and the dimensionless position of the break $m_1$. The distribution \eqref{eq:density-dpl} has a quadratic dependency on the coordinates near the origin:
\begin{equation}
  \hat{\rho}_0 \approx 1 - \frac{k_1}{2} \left( \frac{x^2 + y^2}{a^2} + \frac{z^2}{b^2} \right)  \;.
\end{equation}
Hence, this is characterized by the flat core on the scales of order of the semi-axes scales. At a large distance from the origin, both distributions \eqref{eq:density-spl} and \eqref{eq:density-dpl} are power-laws.

In the two-component model, when the GS/E contribution is taken into account, the total density is:
\begin{equation}
  \label{eq:density-total}
  \hat{\rho}
  = (1 - \xi)\,\hat{\rho}_0 + \xi \hat{\rho}_\mathrm{GS} 
\end{equation}

Here, $\hat{\rho}_0$ represents the density profile given by \eqref{eq:density-dpl}, $\hat{\rho}_\mathrm{GS}$ denotes the GS/E density, and $\xi$ is its relative contribution to the total density. The total density depends on the coordinates $x$, $y$, $z$ and the adjustable parameters $a$, $b$, $k_1$, $m_1$, $k_2$, and $\xi$. As mentioned above, the geometrical parameters of the GS/E profile are fixed to that specified by \cite{Lane2023MNRAS.526.1209L}. Note that the quantity \eqref{eq:density-total} is not normalized as a PDF. For the normalisation, we have to choose the domain volume $V$, then the PDF inside the volume is
\begin{equation}
  \rho = C \hat{\rho}
  \;,\qquad
  C^{-1} = \int_V d^3r\,\hat{\rho}  \;.
\end{equation}
The relative mass fraction of either density component may be estimated as follows:
\begin{equation}
  \label{eq:mass-fraction}
  \xi_0 = C\,(1 - \xi) \int_V d^3r\,\hat{\rho}_0
  \;,\qquad
  \xi_\mathrm{GS} = C \xi \int_V d^3r\,\hat{\rho}_\mathrm{GS}  \;.
\end{equation}

Consider a theoretical model for three-dimensional density distribution $\rho(\vec{r} |\,\theta)$, i.e. the probability density function (PDF) in the Galactocentric position space, where $\theta$ are parameter set of the model. The model density profile should be subject to the same selection effects as the observational data and the sample. Several data selections described in the Sec.~\ref{sec:sample-selection} rely on the star's apparent magnitude. To account for these selection effects, we incorporate the luminosity function in the model, that is, a PDF for the absolute magnitude $\mathrm{M}_\mathrm{G}$, denoted as $\Phi(\mathrm{M}_\mathrm{G})$.

\begin{figure}
  \centering
  \includegraphics[width=0.95\columnwidth]{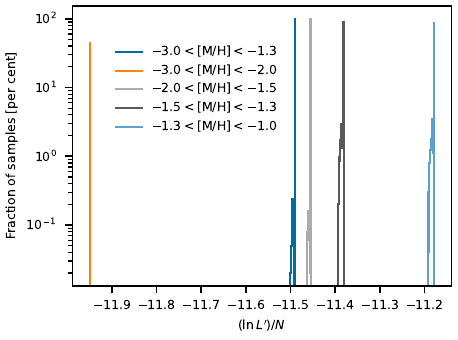}
  \caption{Distributions of the relative log-likelihood values \eqref{eq:loglike} (see details in the Sec.~\ref{sec:spl-nogse}) obtained during a likelihood maximization procedure. Results are for the one-component model \eqref{eq:density-spl} for different values of metallicity.}
  \label{fig:model-convergence}
\end{figure}

Transformation of the theoretical distribution into its observational counterpart inevitably involves some uncertainties due to measurement errors, hence can be considered as a probabilistic transformation \citep[e.g.][]{Boubert2022MNRAS.510.4626B}:
\begin{equation}
  \label{eq:pdf-transformation}
  f(\vec{q}, \mathrm{G} |\,\theta)
  = \int_V d^3r \int d\mathrm{M}_\mathrm{G}\,T(\vec{q}, \mathrm{G}\,|\,\vec{r}, \mathrm{M}_\mathrm{G})
    \,\rho(\vec{r}\,|\,\theta)\,\Phi(\mathrm{M}_\mathrm{G})  \;,
\end{equation}
where $f$ is the joint PDF for the observables, and $T$ is the transformation probability, i.e. the conditional probability (density) to have the observables $(\vec{q}, \mathrm{G})$ given the model quantities $(\vec{r}, \mathrm{M}_\mathrm{G})$. Hereinafter we will refer to $\vec{q} = (l, b, D)$ as the observed position of a star, where $l$ and $b$ are Galactic latitude and longitude, and $D$ is the heliocentric distance.

The observed parallaxes are provided with uncertainties which systematically depend on the celestial coordinates (through the scanning law) and apparent magnitudes \citep{Everall2021MNRAS.502.1908E}, hence will affect the density estimate. These uncertainties can be quite large and can not be neglected, especially for stars in the central parts of the Galaxy. The key idea we implement, is to incorporate the parallax errors into the transformation \eqref{eq:pdf-transformation}, so that the joint PDF $\rho(\vec{r} |\,\theta)\,\Phi(\mathrm{M}_\mathrm{G})$ is smeared along the lines of sight according to the parallax uncertainties. Details of the transformation probability calculation are given in the Appendix \ref{sec:transformation}.

\subsection{Selection function}

To correctly compare the model with the observations, the model should be subjected to the same observational constraints as the data. These constraints are defined as the selection function $S(\vec{q}, \mathrm{G})$, which is the conditional probability for a star to satisfy a set of selection conditions $\mathcal{S}$:
\begin{equation}
  \label{eq:selfun-general}
  S(\vec{q}, \mathrm{G})
  \equiv \mathbb{P}(\mathcal{S} |\,\vec{q}, \mathrm{G})  \;.
\end{equation}
The PDF for the observables after the selection constraints, $f_\mathcal{S}$, can be obtained using the total probability formula:
\begin{equation}
  \label{eq:joint-pdf-constrained}
  f_\mathcal{S}(\vec{q}, \mathrm{G}\,|\,\theta)
  = \frac{1}{Z(\theta)}\,S(\vec{q}, \mathrm{G})\,f(\vec{q}, \mathrm{G}\,|\,\theta)  \;,
\end{equation}
where the denominator $Z$ is the unconditional (except $\theta$) probability to satisfy the constraints,
\begin{equation}
  \label{eq:prb-constraints}
  Z(\theta)
  \equiv \mathbb{P}(\mathcal{S} |\,\theta)
  = \int_V d^3q \int d\mathrm{G}\,S(\vec{q}, \mathrm{G})\,f(\vec{q}, \mathrm{G}\,|\,\theta)  \;.
\end{equation}
The integration here is taken over the spatial domain and the apparent magnitudes range.

\cite{Andrae2023ApJS..267....8A} obtained a relatively pure sample of RGB stars by applying additional constraints designed to reduce contamination, particularly in the low-metallicity regime, from hotter and/or dust-reddened stars. These constraints included a selection by apparent magnitude, parallax, $T_\mathrm{eff}$, $\log_{10}(g)$, and CatWISE colours. Since the model we are developing does not take temperature, surface gravity, or colours into account, these particular cuts could not be incorporated in the model. Therefore, only two selection cuts were inherited from the methodology of \cite{Andrae2023ApJS..267....8A}, i.e. that for the apparent magnitude,
\begin{equation}
  \label{eq:mag-selection}
  S_\mathrm{mag} = \Theta(\mathrm{G} < 16)  \;,
\end{equation}
and for the parallax quality,
\begin{equation}
  \label{eq:plx-selection}
  S_\mathrm{plx} = \Theta(\varpi > 4 \sigma_\varpi)  \;,
\end{equation}
where $\Theta$ is unity if the condition in parentheses is satisfied, and zero otherwise; $\varpi$ and $\sigma_\varpi$ are the parallax and its error, correspondingly. An additional selection was applied to mimic the sampling constraints of \cite{Andrae2023ApJS..267....8A} RGB catalogue: the selection function $S_\mathrm{xp} = S_\mathrm{xp}(l, b, \mathrm{G})$ was defined as a probability that a star in \textit{Gaia} DR3 located at $(l, b, \mathrm{G})$ has an XP spectrum published. The total selection function of the model is the product of the three mentioned above. See the Appendix \ref{sec:parallax-error} and \ref{sec:xp-selection} for full technical details of the selection function calculation.

To apply the selection functions $S_\mathrm{mag}$ and $S_\mathrm{plx}$, the absolute magnitudes need to be converted into the apparent magnitudes. To do this, appropriate amount of extinction must be added to the model magnitudes. The extinction model we use is based on the dust sky map of \citep{Delchambre2023A&A...674A..31D, Green2018JOSS} and on the polynomial approximation of the reddening-temperature dependence \citep{Fitzpatrick2019ApJ...886..108F}, see Sec.~\ref{sec:sample-selection}.

The third panel in the top row of Figure \ref{fig:data-magnitudes} shows the distribution of the extinction $A_\mathrm{G}$ as a function of monochromatic extinction $A_0$ \citep{Fitzpatrick1999PASP..111...63F, Delchambre2023A&A...674A..31D}. The cause of a spread of this function's values is the variation in temperature of the selected stars. Given the selection criteria (stars with a reasonably narrow range of temperatures, see \citet{Andrae2023ApJS..267....8A}) the resulting spread in $A_0$ is quite moderate, not greater than $0.5\,$mag. Since the effective temperature does not enter our model, we only use single-value polynomial estimate $A_\mathrm{G}(A_0)$ (green solid line on the top-right plot in Figure \ref{fig:data-magnitudes}) to compute apparent magnitudes.

\subsection{Parameters optimization}

In order to construct a likelihood function, let us note at first that we focus on the three-dimensional spatial distribution of stars, hence the joint PDF \eqref{eq:joint-pdf-constrained} should be marginalized over the apparent magnitude. Second, it is convenient to bin the observational data onto a grid in the spatial domain, and also to discretise the model over this grid:
\begin{equation}
  \label{eq:pmf}
  p_{jk}(\theta)
  \equiv \int_{\Delta \Omega_j} d\Omega \int_{\Delta D_k} dD\,D^2 \int d\mathrm{G}
    \,f_\mathcal{S}(\vec{q}, \mathrm{G}\,|\,\theta)  \;.
\end{equation}
The volume integral is factorized with two integrals: that over the solid angle $d\Omega$ and that over the distance $dD$. The quantity \eqref{eq:pmf} is a Probability Function (PF) for a selected star to be in a spatial cell $\Delta \Omega_j \times \Delta D_k$. By construction, the PF is normalized in the whole spatial domain of interest:
\begin{equation}
  \sum_{jk} p_{jk} = 1  \;.
\end{equation}

Assume an observational sample consist of $N$ stars, let each $jk$-th spatial cell contain $n_{jk}$ stars, so that $\sum_{jk} n_{jk} = N$. A distribution for the `occupation number' $n_{jk}$ can be described by the multinomial law:
\begin{equation}
  \label{eq:multinomial}
  \mathbb{P}(\{n_{jk}\}\,|\,\theta)
  = N! \prod_{jk} \frac{\left[ p_{jk}(\theta) \right]^{n_{jk}}}{\!\!\!\!\!\!n_{jk}!}  \;.
\end{equation}

\begin{figure}
  \centering
  \includegraphics[width=0.42\textwidth]{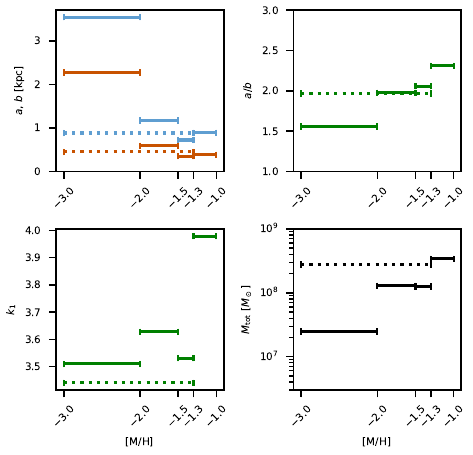}
  \caption{%
  Summary of the one-component model for the metallicity bins from \autoref{tab:summary-nogse}. {\bf Top left:} Semi-axes $a$ (blue) and $b$ (orange). {\bf Top right:} Flattening ratio $a/b$. {\bf Bottom left:} Power-law index $k_1$. {\bf Bottom right:} Total stellar mass. Dotted lines show the values obtained for the best-fit model in the wide metallicity bin.
  }
  \label{fig:model-summary-nogse}
\end{figure}

The problem of optimization of the $\theta$ parameter set to fit the observed distribution $\{n_{jk}\}$ can be formulated as a problem of maximization of a multinomial log-likelihood,
\begin{equation}
  \label{eq:loglike}
  \ln L(\theta)
  \equiv \ln \mathbb{P}(\{n_{jk}\}\,|\,\theta)
  = \sum_{jk} n_{jk} \ln p_{jk}(\theta) + \operatorname{const}  \;,
\end{equation}
where the `$\operatorname{const}$' is independent of $\theta$.

To compare the performance of the statistical model described above with different parameters and samples, it is useful to modify the expression \eqref{eq:loglike}. Specifically, we omit the last term in \eqref{eq:loglike} and then divide the result by the sample size $N$. Hereafter, we denote this as $(\ln L')/N$. In this formulation, maximizing the log-likelihood \eqref{eq:loglike} is equivalent to minimizing the \citet[hereinafter KL]{KullbackLeibler1951} divergence between the model and the sample. Let us denote $q_{jk} = n_{jk}/N$, the maximum likelihood estimate for the observed probability of finding a star in the $jk$-th spatial cell. The KL divergence between the observed and theoretical distributions is
\begin{multline}
  D_\mathrm{KL}[q || p]
  = \sum_{jk} q_{jk}\,\ln\frac{q_{jk}}{p_{jk}(\theta)}  \\
  = \sum_{jk} q_{jk}\,\ln q_{jk} - (\ln L')/N  \;.
\end{multline}
The first term on the right-hand side is independent of the parameter set $\theta$, and can therefore be omitted in the optimization procedure.

To estimate the distribution \eqref{eq:pdf-transformation}, we require a predefined absolute magnitude luminosity function $\Phi(\mathrm{M}_\mathrm{G})$. Our experiments showed that using the observed absolute magnitude luminosity function (LF) of the selected low-metallicity RGB stars (as shown in the histogram for $\mathrm{M}_\mathrm{G}$ in Figure \ref{fig:data-magnitudes}) leads to an overprediction of distant stars (see the histogram for $D$ in the same figure). This occurs because the observed LF is depopulated at the faint end due to selection effects, resulting in an overpopulation at the bright end. Consequently, too many stars in this model satisfy the apparent magnitude condition, $\mathrm{G} < 16$, while being located at large distances from the Sun. 

Instead of using the LF from the RGB star subsample, we adopted the LF from \textit{Gaia} observations of the globular cluster NGC~6397 (represented by the blue thick line in the right panel of Figure \ref{fig:ngc-6397}). For computational accuracy, this LF was smoothed using a piecewise-linear interpolation over 160 points in the absolute magnitude range $-3.5 < \mathrm{M}_\mathrm{G} < 2.5$.

In the next two sections, we apply one- and two-component models to the selected sample of RGB stars. The catalogue of RGB stars of \cite{Andrae2023ApJS..267....8A} is limited to metallicity range $-3.0 < [\mathrm{M}/\mathrm{H}] < -1.0$, and then split into smaller metallicity bins at the thresholds $-2.0$, $-1.5$, and $-1.3$. For each metallicity subsample, we test all possible combinations of the density profile models, \eqref{eq:density-spl} or \eqref{eq:density-dpl}, and combine these with various GS/E model density distributions from \citet{Lane2023MNRAS.526.1209L}.

\begin{figure*}
  \centering
  \includegraphics[width=\textwidth]{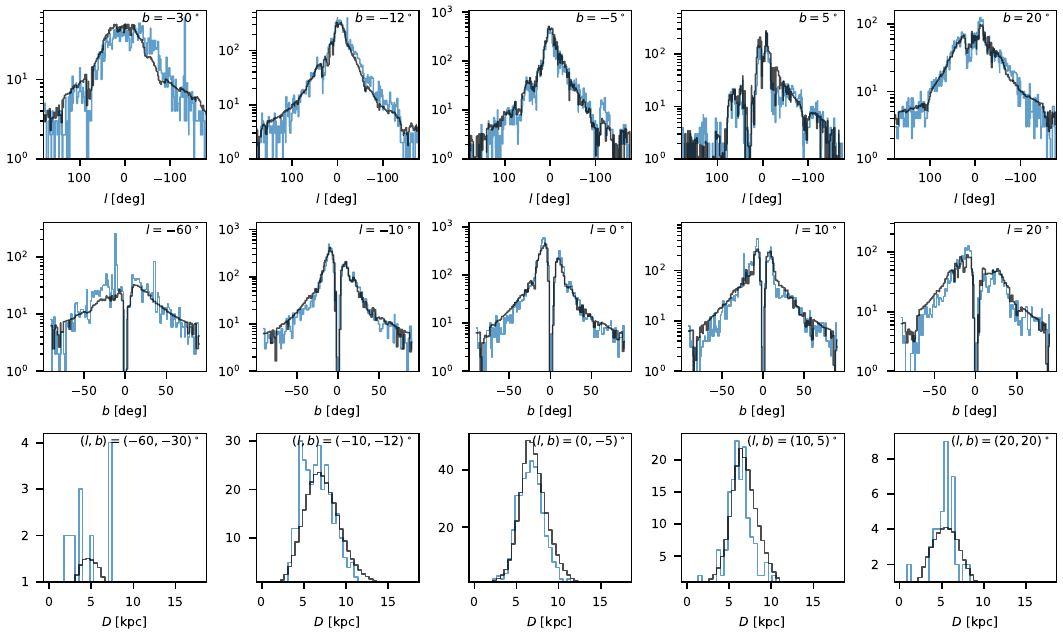}
  \caption{Data vs model comparison in 1D slices of the coordinate space $(l,b,D)$. Distributions of the observed numbers of stars in the metallicity range $-3.0 < [\mathrm{M}/\mathrm{H}] < -1.3$: are shown in blue, while the corresponding model predictions are in black. {\bf Top row:} Constant latitude slices of the sky, i.e. distributions of star counts as a function of Galactic longitude $l$ for various $b$. {\bf Middle row} Constant longitude slices of the sky. {\bf Bottom row:} Distributions of the number counts of stars along the line of sight (see the legends).}
  \label{fig:model-cuts-sharp-nogse-0}
\end{figure*}

\section{One-component model}
\label{sec:spl-nogse}

In this section, we present the results of a single-component model (Eq. \ref{eq:density-spl}) without explicitly accounting for the GS/E's contribution. The computational domain is defined as follows: sky is pixelized with the HEALPix level 5 nested scheme ($12288$ pixels in total); the Galactic distances are binned uniformly into $41$ cells from $0.01$~kpc to $18$~kpc. The transformation between $(l, b, D)$ and $(x, y, z)$ coordinates is performed assuming the Cartesian position of the Sun in the Galaxy is $(-8.12, 0.0, 0.02)$ kpc \citep{GRAVITY2018A&A...615L..15G, Bennett2019MNRAS.482.1417B}.

\begin{table*}
  \makegapedcells
  \setcellgapes{2pt}
  \caption{Summary results of the fitting of the profile \eqref{eq:density-spl} for a set of metallicity bins. The first column is the relative log-likelihood (see the Sec.~\ref{sec:spl-nogse} for details) The $a$, $b$ and $k_1$ columns also contain uncertainties estimated with the Fisher information matrix \eqref{eq:fisher}. The `Sampled' and `Estimated' columns are the numbers of stars in the sample and by the estimate \eqref{eq:nrgb-mod}, correspondingly. The total mass is obtained by Eq.~\eqref{eq:mass-tot}.}
  \centering
  \begin{tabular}{c|ccccccccc}
    \hline
    Metallicity bin
    & \makecell{Log-likelihood\\$(\ln L')/N$}
    & \makecell{X-Y scale\\$a$ [kpc]}
    & \makecell{Z scale\\$b$ [kpc]}
    & \makecell{Power index\\$k_1$}
    & \makecell{Flattening\\$a/b$}
    & \makecell{Sampled\\$N$}
    & \makecell{Estimated\\$N_\mathrm{RGB}$}
    & \makecell{Total mass\\$M_\mathrm{tot}$ [$M_\odot$]}
    \\
    \hline
    \hline
    & \multicolumn{7}{c}{\textit{NGC~6397 LF}}
    \\[-4pt]
    \makecell[t{{c}}]{%
        $-3.0 \;\cdots\, -1.3$ \\[2pt]
        $-3.0 \;\cdots\, -2.0$ \\[2pt]
        $-2.0 \;\cdots\, -1.5$ \\[2pt]
        $-1.5 \;\cdots\, -1.3$ \\[2pt]
        $-1.3 \;\cdots\, -1.0$
    }
    & \makecell[t{{r}}]{$-11.489$ \\[2pt] $-11.949$ \\[2pt] $-11.455$ \\[2pt] $-11.381$ \\[2pt] $-11.178$}
    & \makecell[t{{r}}]{$0.88\pm0.03$ \\[2pt] $3.54\pm0.09$ \\[2pt] $1.17\pm0.04$ \\[2pt] $0.72\pm0.05$ \\[2pt] $0.89\pm0.03$}
    & \makecell[t{{r}}]{$0.45\pm0.03$ \\[2pt] $2.28\pm0.09$ \\[2pt] $0.59\pm0.04$ \\[2pt] $0.35\pm0.05$ \\[2pt] $0.39\pm0.03$}
    & \makecell[t{{r}}]{$3.441\pm0.005$ \\[2pt] $3.510\pm0.033$ \\[2pt] $3.628\pm0.008$ \\[2pt] $3.530\pm0.006$ \\[2pt] $3.980\pm0.004$}
    & \makecell[t{{r}}]{$1.97$ \\[2pt] $1.56$ \\[2pt] $1.98$ \\[2pt] $2.06$ \\[2pt] $2.31$}
    & \makecell[t{{r}}]{$219\,325$ \\[2pt] $23\,462$ \\[2pt] $105\,109$ \\[2pt] $90\,754$ \\[2pt] $214\,591$}
    & \makecell[t{{r}}]{$3\,293\,529$ \\[2pt] $482\,659$ \\[2pt] $1\,326\,066$ \\[2pt] $1\,278\,025$ \\[2pt] $2\,796\,238$}
    & \makecell[t{{r}}]{$2.76\times10^8$ \\[2pt] $2.48\times10^7$ \\[2pt] $1.28\times10^8$ \\[2pt] $1.25\times10^8$ \\[2pt] $3.39\times10^8$}
    \\
    \hline
    & \multicolumn{7}{c}{\textit{MIST LF}}
    \\[-4pt]
    $-3.0 \;\cdots\, -1.3$
    & $-11.504$ & $0.54\pm0.04$ & $0.29\pm0.04$ & $3.342\pm0.004$ & $1.87$ & $219\,325$ & $4\,615\,145$ & $3.01\times10^8$
    \\
    \hline
  \end{tabular}
  \label{tab:summary-nogse}
\end{table*}

Searching for an optimal set of parameters is carried out by solving a minimization problem for the negative log-likelihood \eqref{eq:loglike} as a function of $\theta = [b, a/b, k_1]$. The reason for not using a convenient posterior-based inference method such as Markov chain Monte Carlo is that the domain consists of more than $500\,000$ cells (level 5 HEALPixels $\times$ distance grid). As a result, the multinomial probability \eqref{eq:multinomial} takes very small values on average with an extremely high contrast in a very narrow region around the optimal parameters. In practice, we use the `\textsc{L-BFGS-B}' method\footnote{%
The whole model is implemented on \textsc{Python~3.11} \citep{vanRossum2009doi10.5555/1593511} + \textsc{Numpy} \citep{Harris2020array} + \textsc{Scipy} \citep{2020SciPy-NMeth} + \textsc{healpy} \citep{Zonca2019, Gorski2005ApJ...622..759G} + \textsc{Astropy} \citep{astropy:2013, astropy:2018, astropy:2022} + \textsc{sqlutilpy} \citep{Koposov2024Zenodo6396777}.}%
(this is an extension of the Limited memory Broyden–Fletcher–Goldfarb–Shanno algorithm that allows search bounds to be specified, \citet{Liu1989MathProg}) for optimizing the likelihood function \eqref{eq:loglike}. To avoid sub-optimal performance due to potential trapping in the local minima, the best log-likelihood was chosen among $10\,000$ independent runs with the initial values chosen randomly and uniformly in the following limits:
\begin{gather}
  0.1 \leqslant b \leqslant 3.0  \;,  \\
  1.0 \leqslant a/b \leqslant 2.5  \;,  \\
  3.0 \leqslant k_1 \leqslant 7.0  \;.
\end{gather}

Each run successfully found the maximum of the likelihood function \eqref{eq:loglike}. While the optimization algorithm is not perfect and may not always converge to the same solution, it consistently converged to the same point for each model in the vast majority of cases. The convergence success rates are shown at Figure \ref{fig:model-convergence}. As seen, the optimization routine converged to the same point in most of the runs for each metallicity bin. In the case of the lowest metallicity, the performance is slightly weaker. Note, that the log-likelihood values at a horizontal axis of the Figure \ref{fig:model-convergence} differ from \eqref{eq:loglike} in that there is no constant term in them, and they are also related to the number of stars in the corresponding sample (see the explanations in the previous Section). The latter is done so that it is possible to compare the performance of the model for different metallicity samples, and hence different numbers of stars in the samples.

Table~\ref{tab:summary-nogse} and Figure \ref{fig:model-summary-nogse} give the summary of the best-fitting results. The stellar population of the lowest metallicity subsample has a less flattened distribution compared to the highest-metallicity one. The latter is also much more concentrated (the axes are smaller and the density drop is steeper). For metallicity $<-1.3$, there is an apparent trend in the axes size and the flattening value. On the other hand, the trend in the power index is much weaker, with $k_1 \approx 3.5$. By number, the combined population in the wide metallicity bin, $-3.0 < [\mathrm{M}/\mathrm{H}] \leq -1.3$, is dominated by the stars with higher metallicities, so the scales, flattening, and the power index values tend to the values in the high(er) metallicity bin(s).

With the best-fit values of the model parameters $\theta_\ast$ in hand, we can estimate the uncertainties in $a$, $b$ and $k_1$ using the Fisher information matrix for the multinomial distribution:
\begin{equation}
  \label{eq:fisher}
  I_{\alpha\beta}
  = N \sum_{jk} \frac{1}{p_{jk}}\,\pdiff{p_{jk}}{\theta_\alpha}\,\pdiff{p_{jk}}{\theta_\beta}
    \biggr|_{\theta = \theta_\ast}  \;,
\end{equation}
where $\partial/\partial \theta_\alpha$ is the $\alpha$-th parameter derivative. The square root of the diagonal of the inverse of this matrix gives the uncertainties of the corresponding parameter. For all of the metallicity bins, the uncertainty in the semi-axes did not exceed $14$ per cent, and the power index uncertainty was less than $\sim2$ percent (see \autoref{tab:summary-nogse}).

Let us estimate the total stellar mass of the Galaxy component represented by the RGB stars in our sample. Given the number of stars in the sample $N$ and the best fitting parameters $\theta_\ast$, it is possible to estimate a total number of RGB stars in the sample, $N_\mathrm{RGB}$, \textit{as if no selection constraints were applied}:
\begin{equation}
  \label{eq:nrgb-mod}
  N_\mathrm{RGB}
  = \frac{N}{Z(\theta_\ast)}  \;,
\end{equation}
where $Z$ is the probability to fulfil the selection constraints, Eq.~\eqref{eq:prb-constraints}. Indeed, the probability to satisfy the selection criteria is essentially a fraction of a true number of stars that passed the constraints and were observed. Since the sample we use contains only the RGB stars, the estimate \eqref{eq:nrgb-mod} is a number of RGB stars in the population. On the other hand, consider an isochrone of a given age and metallicity. Let $N_\mathrm{tot}$ be a total number of all stars in the population, $m_\mathrm{lo}$ and $m_\mathrm{hi}$ are the lowest and the highest stellar masses on RGB branch of the isochrone, correspondingly. Then the number of RGB stars is
\begin{equation}
  \label{eq:nrgb-iso}
  N_\mathrm{RGB}
  = N_\mathrm{tot} \int_{m_\mathrm{lo}}^{m_\mathrm{hi}} dm\,\phi(m)  \;,
\end{equation}
where $\phi$ is an initial mass function (IMF). Using the IMF, we also can estimate the total mass of the entire population:
\begin{equation}
  \label{eq:mass-tot}
  M_\mathrm{tot}
  = N_\mathrm{tot} \int_{m_\mathrm{min}}^{m_\mathrm{hi}} dm\,m \phi(m)  \;,
\end{equation}
where $m_\mathrm{min}$ is the minimum stellar mass on the isochrone. Assuming that (currently) the most massive stars in the population are the RGBs, the Eq.~\eqref{eq:mass-tot} gives the mass estimate for the entire population\footnote{%
This method is similar to that used by \cite{Mackereth2020MNRAS.492.3631M}.}. The total number of stars $N_\mathrm{tot}$ is obtained with \eqref{eq:nrgb-iso} and \eqref{eq:nrgb-mod}. We use Kroupa IMF \citep{Kroupa2001MNRAS.322..231K} with $m_\mathrm{min} = 0.1\,M_\odot$ and MIST isochrones \citep{Choi2016ApJ...823..102C} accessed via \textsc{minimint} Python interface \citep{Koposov2023Zenodo4002971}. The $M_\mathrm{tot}/N_\mathrm{RGB}$ relations for populations with an age of $13\,$Gyr and various metallicities are given at the \autoref{tab:iso}. These values are used to compute $M_\mathrm{tot}$ in the \autoref{tab:summary-nogse}. As can be seen, one RGB star accounts approximately for $140\;M_\odot$ of the mass of the entire stellar population.

The $M_\mathrm{tot}/N_\mathrm{RGB}$ ratio can also be directly estimated from observations of globular clusters. Specifically, we used the data for two GCs: NGC~6121 ($D = 1.8\;$kpc, $[\mathrm{Fe}/\mathrm{H}] = -1.18$) and NGC~6397 ($D = 2.5\;$kpc, $[\mathrm{Fe}/\mathrm{H}] = -1.99$) from the catalogs of \citet{Vasiliev2021Zenodo4891252, Vasiliev2021MNRAS.505.5978V} and \citet{Baumgardt2023MNRAS.521.3991B}. For both clusters, a color-absolute magnitude diagram was constructed, and the RGB stars were selected. As these clusters are close to the Sun, the RGB stars are not subject to strong selection effects. The ratio of the dynamical mass of the NGC~6121 cluster to the number of its RGB stars was found to be approximately $130\;M_\odot$. For NGC~6397, this value is about $135\;M_\odot$. These estimates are in good agreement with the value obtained earlier using theoretical isochrones.

It is nice to note that the model turns out to be additive in the sense that the estimate of the total mass of the stellar population obtained for the wide metallicity bin coincides with the sum of estimates obtained for the narrower bins (\autoref{tab:summary-nogse}). The total mass in the whole interval of metallicities from $-3$ to $-1$ turns out to be $\approx 6\times10^8\:M_\odot$.

\begin{figure*}
  \centering
  \includegraphics[width=\textwidth]{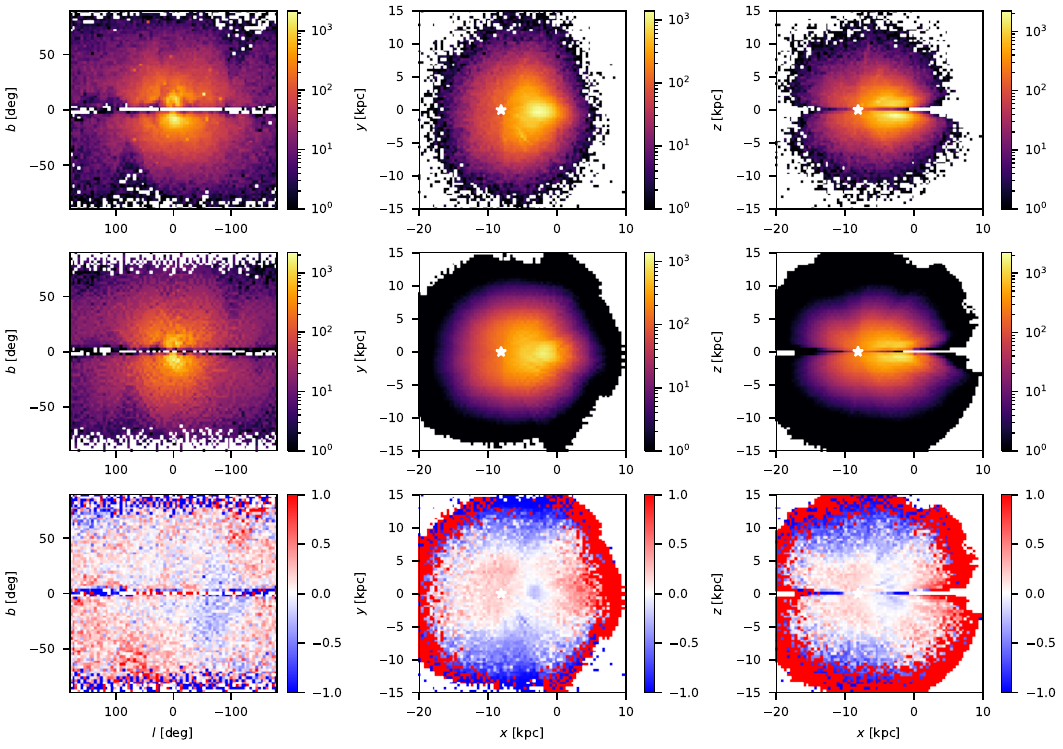}
  \caption{Comparison of 2D distributions of the number counts of stars with $-3.0 < [\mathrm{M}/\mathrm{H}] < -1.3$ in the data ({\bf top}), the model ({\bf middle}) and their scaled difference $(\mathtt{model} - \mathtt{data})\,/\,(\mathtt{model} + \mathtt{data})$ ({\bf bottom}). {\bf Left column} Projection of the stellar density on the sky in Galactic coordinates. Small strong over-densities are Galactic globular clusters. The residuals are well-behaved except for moderate over-predictions in regions strongly affected by the {\it Gaia} selection function (e.g. at $l\approx80^{\circ}, b\approx-50^{\circ}$). There is also a broad region at $l\approx-80^{\circ}, b\approx0^{\circ}$ where the model under-predicts the counts slightly, this may be connected to the presence of the so-called Virgo Overdensity \citep[and references therein]{Simion2019MNRAS.482..921S} and/or the Magellanic system. {\bf Centre column:} Projections of the stellar density onto the $X-Y$ plane in Galactocentric Cartesian coordinates. Here a stronger blue-red residual pattern is observed, possibly indicating a departure from the axi-symmetry in the observed counts. {\bf Right column:} Projections of the stellar density onto the $X-Z$ plane in Galactocentric Cartesian coordinates. White star marks the position of the Sun.}
  \label{fig:model-proj-sharp-nogse-0}
\end{figure*}

\begin{table}
  \makegapedcells
  \setcellgapes{3.0pt}
  \caption{See details in Sec.~\ref{sec:spl-nogse}.}
  \centering
  \begin{tabular}{c|ccccc}
    \hline
    $[\mathrm{M}/\mathrm{H}]$
    & $-3.0$ & $-2.0$ & $-1.5$ & $-1.3$ & $-1.0$
    \\
    \hline
    $M_\mathrm{tot}/N_\mathrm{RGB}$
    & $144.7$ & $142.6$ & $140.5$ & $140.3$ & $133.9$
    \\
    \hline
  \end{tabular}
  \label{tab:iso}
\end{table}

Figure \ref{fig:model-cuts-sharp-nogse-0} compares the data (blue) and the best-fit model (black) for different locations in the Galaxy. As the Figure demonstrates, in most $b, l$ bins, the stellar density distribution is strongly jagged. This is due to the XP selection function. Additionally, it is possible that some of the narrow peaks may be caused by globular clusters (cf. Figure \ref{fig:data-obs-sky}).  The broader peaks and troughs are is due to the joint action of the selection in parallax and in apparent magnitudes.
Notwithstanding the small sub-structure in the density slices, overall, the observed density variations are captured well by the model.

Figure~\ref{fig:model-proj-sharp-nogse-0} shows projections of the the observed star counts (top), the model (middle) and the scaled residuals (bottom) in the heliocentric Galactic celestial coordinates (left) and in the principal planes of the Galacto-centric Cartesian coordinate system (centre and right). The `triangles' of low visibility are also seen in the $l$-$b$ projection of the model (middle left), although they are not as deep as in the observed data (top left). This slight under-performance of the model selection function is also reflected in the residuals shown in the bottom left panel. Similar to Figure \ref{fig:model-cuts-sharp-nogse-0}, the selection cuts in apparent magnitude and parallax lead to the asymmetry of the distribution of stars relative to the Galactic centre along the $X$ axis. Small strong over-densities in the top left panel are Galactic globular clusters. There is also a broad region at $l\approx-80^{\circ}, b\approx0^{\circ}$ where the model under-predicts the counts slightly, this may be connected to the presence of the so-called Virgo Overdensity \citep[and references therein]{Simion2019MNRAS.482..921S} and/or the Magellanic system.

Figure \ref{fig:model-radial-sharp-nogse-0} shows the Galactocentric radial profiles of the observed volume number density of stars (blue), the best-fit model with the selection function applied (black solid) together with two variations of the model with two slightly different power-law indexes, as well as the best-fit model unaffected by the selection (black dashed), which is compared to the distribution of the MW Globular Clusters (green histogram), which we discuss further below. As the Figure demonstrates, once the selection is applied, within $\approx 12$~kpc from the Galactic centre, the model fits the data reasonably well. This is also illustrated by a flat and low-amplitude residual curve shown in the bottom panel. Beyond $\approx 12$ kpc, the observed star counts exceed the model. There is also a data-model mismatch in the inner $\approx 1.5$ kpc, this region is strongly affected by the effects of the selection exacerbated by the dust absorption. A possible underestimation of the selection strength in the central area may be a reason for some overestimation of the model density at $\approx 1.5$~kpc radius. The selection underestimation is also noticeable in the $l$-$b$ projection on the Figure \ref{fig:model-proj-sharp-nogse-0}. Radial profiles for other metallicity bins are shown on Figure \ref{fig:model-radial-sharp-nogse-1-4}.

\begin{figure}
  \centering
  \includegraphics[width=0.98\columnwidth]{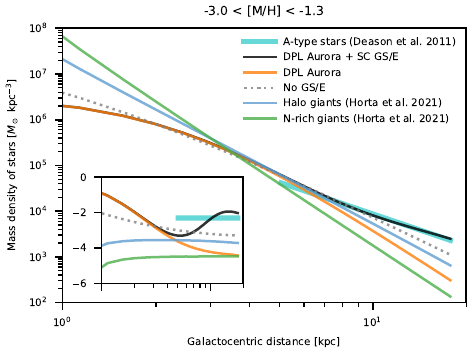}
  \caption{Mass density distributions for double-component model `DPL Aurora \& SC GS/E' ($-3.0 < [\mathrm{M}/\mathrm{H}] < -1.3$), model with no GS/E, model by \citet{Horta2021MNRAS.500.5462H} of the distribution of low-metallicity giants ($-2.5 < [\mathrm{M}/\mathrm{H}] < -1.0$) and of N-rich stars (multiplied by $10$) in the same metallicity range, and the model for A-stars distribution \citep{Deason2011} (scaled arbitrarily). The inset shows the log-log derivative (i.e. the local slope) of the corresponding density distributions.}
  \label{fig:model-radial-horta}
\end{figure}

Our results can be compared to those reported by \cite{Horta2021MNRAS.500.5462H} who modelled the density distribution of the low-metallicity halo stars in the APOGEE DR16 data \citep{Abdurrouf2022ApJS..259...35A}. For their triaxial power-law density model they report the power-law index of $3.48$, which is in perfect agreement with our results ($3.44$ in the wide metallicity bin, Figure \ref{fig:model-radial-horta}). Note that \cite{Horta2021MNRAS.500.5462H} only modelled the APOGEE data in the range of galactocentric distances from $1.5$~kpc to $15$~kpc. Compared to the model used here, the model of \cite{Horta2021MNRAS.500.5462H} has a cusp that diverges at the centre, hence the two models disagree in the innermost MW. The \cite{Horta2021MNRAS.500.5462H} estimate of the total halo stellar mass in the above-mentioned distance range was $8.3\times10^8\,M_\odot$, a factor of $\approx 2$ higher than ours but the two models agree within $3\sigma$.

As already mentioned, Figure~\ref{fig:model-radial-sharp-nogse-0} also compares the reconstructed model density of RGB stars with $-3.0 < [\mathrm{M}/\mathrm{H}] < -1.3$ (black dashed line) and the observed number density of Globular Clusters (green histogram). The two radial density distributions show remarkable agreement across almost the entire range of Galactocentric distances probed. Only within $r \approx 2$ kpc, the GC profile starts to drop below our RGB model density. This is unsurprising: the census of the Galactic GCs is likely to be incomplete in the very centre. This incompleteness affects most profoundly the low-mass clusters, as illustrated in Figures 7 and 8 of \citet{Baumgardt2019}. With time, induced by Galactic tides, this GC dissolution should indeed lead to a flattening of the cluster density profile. As figures in \citet{Baumgardt2019} demonstrate, this selection effect is most pronounced inside $2$ kpc of the Galactic centre and is less noticeable further out. The match between the GC number density profile and the stellar density is perplexing. One possible explanation is perhaps that a large portion of the stellar mass in this metallicity range in the inner Galaxy is contributed by stars removed from the surviving most massive GCs \citep[see also the discussion in][]{Belokurov2023}.

\begin{figure}
  \centering
  \includegraphics[width=0.49\textwidth]{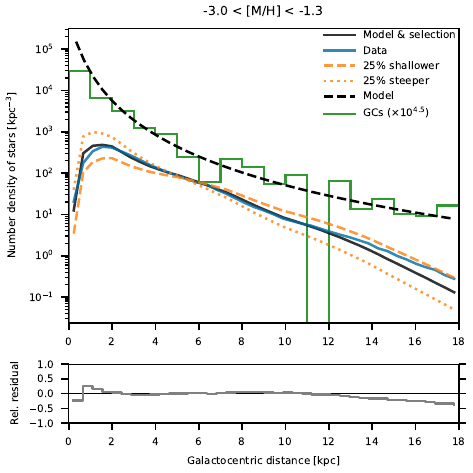}
  \caption{Single-component model of the radial density profile of low-metallicity giants in the centre of the Galaxy. {\bf Top panel:} Observed number density of stars with $-3.0 < [\mathrm{M}/\mathrm{H}] < -1.3$ (solid blue) is compared to the theoretical model \eqref{eq:density-spl} (black solid line) with selection effects applied. For illustration we also plot models with the power-law index varied by  $25$ per cent (orange). The underlying model unaffected by the sample selection is shown as a black dashed line. For comparison, the number density of Galactic globular clusters is plotted in green. {\bf Bottom panel:} Relative residuals between the model and the data: $(\mathtt{model} - \mathtt{data})\,/\,(\mathtt{model} + \mathtt{data})$.}
  \label{fig:model-radial-sharp-nogse-0}
\end{figure}

In additional to using the luminosity function of RGB stars in the globular cluster NGC~6367, we also tested the model with a synthetic LF. The distribution of the magnitudes of RGB stars was generated with the Kroupa IMF \citep{Kroupa2001MNRAS.322..231K} and the MIST isochrone \citep{Choi2016ApJ...823..102C} of an age of $13.3$~Gyr and metallicity $-1.99$ \citep{Baumgardt2023MNRAS.521.3991B}. The results of applying this LF to the wide metallicity bin $-3.0 < [\mathrm{M}/\mathrm{H}] < -1.3$ are reported in Table~\ref{tab:summary-nogse} (the `\textit{MIST LF}' part). As the bottom row of the Table demonstrates, the model with a synthetic LF predicts a more compact and less steep core. The likelihood of this model is worse than the model where the NGC~6367 LF was used. At the same time, the relative difference of the semi-axes' scales between the two models is about $22$ per cent, and the relative difference in the power indexes is about $1.5$ per cent. Thus we conclude that the models are consistent to each other, but the preference should be given to the `\textit{NGC~6367 LF}' model.

\section{Two-component model}
\label{sec:dpl}

As discussed in Section~\ref{sec:intro}, several pieces of observational evidence point to the presence of at least two distinct halo components in the inner Milky Way. As pointed out soon after Gaia Data Release 2 by \citet{Myeong2018gcs}, a certain "critical" orbital energy exists which neatly separates accreted and in-situ globular clusters. Echoing this idea, \citet{Belokurov2023} show that indeed a sharp boundary can be drawn in the energy $E$ and angular momentum $L_z$ space to separate field stars with distinct levels of [Al/Fe], indicative of different rates of early star-formation and self-enrichment. Subsequently, \citet{Belokurov2024gc} demonstrate that this boundary also separates Galactic GCs into in-situ and accreted, in agreement with the earlier result of \citet{Myeong2018gcs}. Curiously, the boundary passes very close to the Solar orbital energy, implying that around the Sun, the two halo components contribute approximately equally. Guided by the results of these studies, in this section, we report the results of the two-component fit. The distribution of the selected RGB stars is described here using a combination of the profile \eqref{eq:density-dpl} and one of the GS/E density models presented in \citet{Lane2023MNRAS.526.1209L}.

\citet{Lane2023MNRAS.526.1209L} identify the likely GS/E stars amongst the APOGEE data using a set of sophisticated kinematic cuts. The GS/E sample is then approximated using the APOGEE selection function and a set of parameterised models. \citet{Lane2023MNRAS.526.1209L}'s GS/E model is a triaxial ellipsoid centered at the Galactic centre. The density distribution is constructed as a combination of power-laws and exponentials, described by several parameters including the orientation of the ellipsoid. \citet{Lane2023MNRAS.526.1209L} experiment with several kinematic selections of the GS/E stars in the APOGEE data, hence their density models are fitted to the different datasets. In our analysis we use the following models of the GS/E: i) an exponentially truncated single power-law after 'action diamond' selection%
\footnote{For variety of the selections, please see \citet{Lane2023MNRAS.526.1209L}.} hereinafter referred to as `SC GS/E'; ii) a broken power-law with exponential truncation after the 'action diamond' selection, referred to as `BPL GS/E'; iii) a double-broken power-law with disc contamination after `$e \mathdash L_z$' selection, referred to as `DBPL+D GS/E'. Our choice of the GS/E density profiles is motivated not only by the rich variety of the functional forms but also by behaviour of the profiles at the origin. \citet{Lane2023MNRAS.526.1209L} have constrained their models in the Galactocentric distance ranging from $2$~kpc to $70$~kpc. In this Paper, we are particularly interested in the Aurora population, hypothesised to be strongly centrally concentrated. To isolate the Aurora population we select stars with low metalicities, i.e. [M/H]$<-1$, and more conservatively, assume that our selection contains little of disc or Splash below [M/H]$=-1.3$ \citep{Belokurov2020MNRAS.494.3880B}. Note that, the characteristic scale of the Aurora population is expected to be of order of few kpcs. With that in mind, the GS/E models of choice must not diverge at $r = 0$.

To find the best-fit two-component model \eqref{eq:density-total}, we fix the geometrical parameters of the GS/E models as given in \citet{Lane2023MNRAS.526.1209L} and vary the parameters of the Aurora density component described by Eq. \eqref{eq:density-dpl}. In this exercise the GS/E component acts as a kind of background. As in Sec.~\ref{sec:spl-nogse}, the fits are done using the iterative optimization procedure, with the trial values of the parameters (including the GS/E contribution to the overall density) are chosen randomly in the wide limits.
The results are summarized in the \autoref{tab:summary-dpl}.

\begin{figure}
  \centering
  \includegraphics[width=0.46\textwidth]{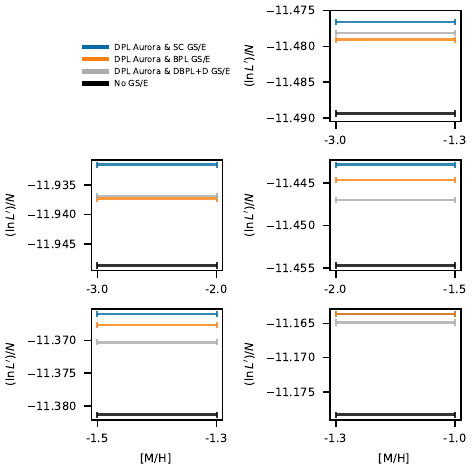}
  \caption{Relative log-likelihood \eqref{eq:loglike} (see Sec.~\ref{sec:statistical-model}) for all types of models considered in the Paper. Single-component model (black) is under-performing significantly compared to its more complex two-component counterparts.}
  \label{fig:model-loglikes}
\end{figure}

In order to compare two-component models with different GS/E parameterisations, the relative log-likelihood values are also included to the \autoref{tab:summary-dpl} (see also Figure \ref{fig:model-loglikes}). As seen, the Aurora model with the `SC' GS/E background better fits the samples (has greater values of the log-likelihood) than any other model among all metallicity bins, while using the `DBPL+D' GS/E background lead to a worse fit. We also not that the one-component (`No GS/E') model has systematically worse likelihood over all metallicity ranges than the two-component model.

Figure \ref{fig:model-loglikes} contains relative log-likelihood values for all models calculated in the present Paper. As we can see, the two-component models with the `SC' profile for GS/E have the highest log-likelihood values in the same metallicity bins, so they should be favoured. It would be useful however to quantify this preference by a statistical criterion. Let us use the Bayes factor for this purpose, a ratio of evidences of two models, where the evidence is an unconditional probability for the model to generate the observed sample:
\begin{equation}
  K
  = \frac{\int d\theta\,\mathbb{P}_\text{A}(\{n_{jk}\}\,|\,\theta)\,Q_\text{A}(\theta)}%
    {\int d\theta\,\mathbb{P}_\text{B}(\{n_{jk}\}\,|\,\theta)\,Q_\text{B}(\theta)}  \;.
\end{equation}
Here $\mathbb{P}_\text{A}(\{n_{jk}\}\,|\,\theta)$ is the likelihood function \eqref{eq:multinomial} for a model A and $Q_\text{A}(\theta)$ is the prior PDF for $\theta$, the parameter set. According to \cite{Kass1995JASA}, the case $K > 10$ is considered as being strongly in favor of the model A, and the ratio $K > 100$ is considered as decisive. In the Tables \ref{tab:summary-nogse} and \ref{tab:summary-dpl}, we have seen that the actual errors of the parameters are $\lesssim 10$ per cent of their optimal values. Using this fact, we may speculate the following approximation for the priors: $Q_\text{A}(\theta) = |I_\text{A}|^{-1} \delta(\theta - \theta_{\ast\text{A}})$, where $|I_\text{A}|$ is the determinant of the Fisher information matrix \eqref{eq:fisher} and $\theta_{\ast\text{A}}$ is the optimal parameter set for the model A. The same is assumed for model B. Given this, the logarithm of the Bayes factor is
\begin{equation}
  \label{eq:log-bayes-rate}
  \ln K
  = \ln L'_\text{A} - \ln L'_\text{B} + \ln\frac{|I_\text{B}|}{|I_\text{A}|}  \;,
\end{equation}
where $\ln L'_\text{A}$ and $\ln L'_\text{B}$ are the log-likelihood value taken from the \autoref{tab:summary-dpl} and \autoref{tab:summary-nogse}, correspondingly (note that the relative log-likelihood values are given in these tables). Suppose the model A is the two-component model with the `SC' background (the model with a highest log-likelihood at the Figure~\ref{fig:model-loglikes}) and the model B is the one-component model, i.e. with no GS/E (the lowest log-likelihood one). In this case the $\ln K = 2816$, which is huge enough to choose the first model, `DPL Aurora \& SC GS/E'.

\begin{figure}
  \centering
  \includegraphics[width=0.42\textwidth]{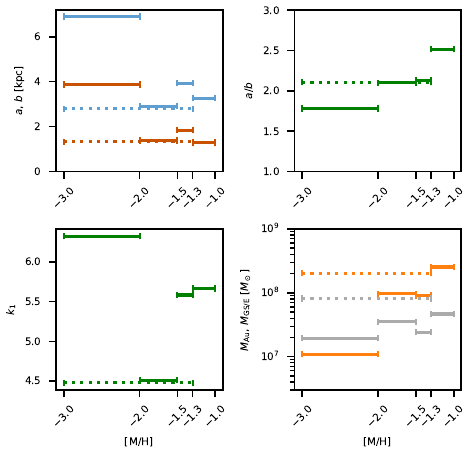}
  \caption{A two-component model with DPL Aurora \& SC GS/E. Similar to Figure~\ref{fig:model-summary-nogse}, this gives  a summary of the best-fit model parameters in various metallicity bins, see \autoref{tab:summary-dpl}. {\bf Top left panel:} horizontal (blue) and vertical (red) scales of the model. {\bf Top right:} Flattening $a/b$. {\bf Bottom left:} Power-law index of the inner slope.  {\bf Bottom right:} Estimated stellar mass of the Aurora (orange) and the GS/E (grey) populations.}
  \label{fig:model-two-comp-params}
\end{figure}

\begin{figure}
  \centering
  \includegraphics[width=0.45\textwidth]{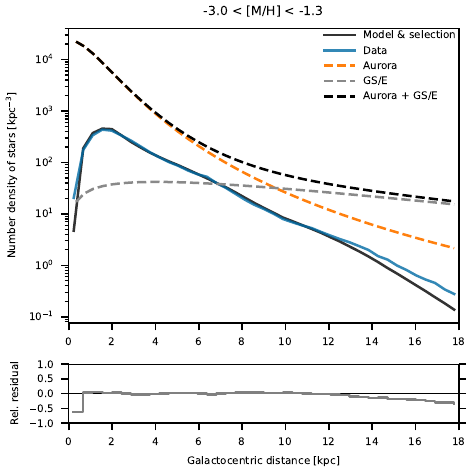}
 \caption{DPL Aurora \& SC GS/E model. Radial density profiles for sample (blue line), for Aurora model (dashed orange line), for GS/E model (dashed grey), for sum of the last two (dashed black), and for total density after selections applied (solid black line). Bottom panel shows the relative residuals between the sample and the model after the selections applied: $(\mathtt{model} - \mathtt{data})\,/\,(\mathtt{model} + \mathtt{data})$.}
  \label{fig:model-two-comp-radial}
\end{figure}

Figure \ref{fig:model-two-comp-params} gives the summary of the best-fit parameters of our preferred two-component model with a DPL Aurora and a SC GS/E for different metallicity ranges. The behaviours of other two-component models can be inspected in  Appendix~\ref{sec:add-figs}. One thing that remains the same irrespective of the number (or type) of components used is the overall vertical flattening of the density: the ratio of the horizontal and vertical scales $a/b\approx2$ in all cases. As Figure \ref{fig:model-two-comp-params} demonstrates, compared to the one-component model, in the two-component model the Aurora's density is steeper but the semi-axes scales are significantly larger, indicating a flattening in the density inside $\approx 2.8$ kpc. Going back to other two-component models, when compared at the same metallicity, the semi-axes scales are quite similar across all GS/E backgrounds tested. However, for any two-component model, the scales are systematically larger than in the one-component model at the corresponding metallicities (see the Sec.~\ref{sec:spl-nogse} and the \autoref{tab:summary-nogse}). The inner power-law indices, $k_1$, are significantly steeper and more diverse than in the models with no GS/E. This is expected and can be explained by the fact that the GS/E profile is much flatter than the hypothetical Aurora, dominating the overall density at large radii. Not taking the GS/E explicitly into account makes the one-component model flatter. The total Aurora mass from the two-component model is $\approx2\times10^8 M_{\odot}$, meaning that this component is responsible for $\approx 2/3$ of the stellar halo in the Galactocentric range considered.

\begin{figure}
  \centering
  \includegraphics[width=0.98\columnwidth]{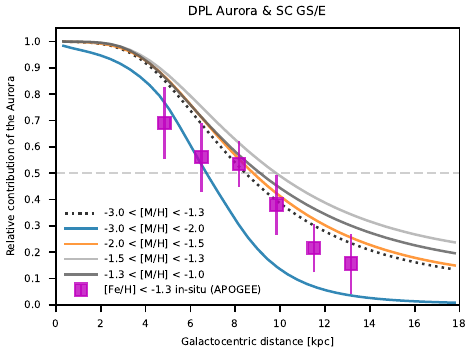}
  \caption{Contribution of the Aurora to the total mass depending on distance to the Galactic centre. The DPL Aurora \& SC GS/E model is shown in various metallicity bins. A level at which the Aurora and GS/E contributions are equal is marked by the horizontal light grey dashed line. The magenta boxes shows the estimate for in-situ stars with metallicities $[\mathrm{Fe}/\mathrm{H}] < -1.3$ obtained using the APOGEE data (see Sec.~\ref{sec:dpl}).}
  \label{fig:model-radial-aurora-flat-sc}
\end{figure}

Figure~\ref{fig:model-two-comp-radial} shows the behaviour of the two-component model discussed above as a function of the Galactocentric radius, analogously to Figure~\ref{fig:model-radial-sharp-nogse-0} which presents the one-component model. Here, the residuals (shown in the bottom panel) are flatter overall, and hence compared to one-component model, the two components do a better job across the entire range of radii. The most striking improvement is inside the inner $\approx$2 kpc, where the one-component model struggled noticeably. The two-component fit residuals remain flat from $r\approx10$ kpc down to our smallest distance bin of $r<0.45$ kpc where the model largely under-predicts the data. We note however, that this particular radial bin has one of the most uncertain star counts due to small number statistics and severe effects of extinction, blending and selection. Even in the outer regions, i.e. for the radial range $10<r$(kpc)$<18$, two components systemically perform better than one, giving $\approx15\%$ drop in the mean residual outside the Solar radius.

In some cases, including that presented in Figures ~\ref{fig:model-two-comp-params} and \ref{fig:model-two-comp-radial}, the unitless break radius $m_1$ exceeded the domain range (these cells are stricken out in the \autoref{tab:summary-dpl}). As a result, the model effectively becomes a single power-law. The second power-law index, $k_2$, when present, shows significant scatter across all GS/E backgrounds. This can be seen especially in the lowest metallicity bin, $-3.0 < [\mathrm{M}/\mathrm{H}] < -2.0$. The reason for this is that at largest radii, the GS/E component dominates over, or is at least comparable to, the Aurora component. Hence, given a fixed GS/E model, the outer radial profile of Aurora is guided by the radial trend in GS/E which is varied from model to model.

Across all models, the lowest-metallicity ($-3.0 < [\mathrm{M}/\mathrm{H}] < -2.0$) population has a wide and very steep core ($k_1 = 6.32 \dots 7.33$). A similar but less pronounced behaviour is also seen in the population with metallicity values $-1.5 \dots -1.3$. In general, the flatness $a/b$ grows as the metallicity increases. This can be interpreted as a transition to the disc morphology.

The masses of the Aurora and GS/E in our models (\autoref{tab:summary-dpl}) were estimated over the Galactocentric radius range of $0.1$ to $18$~kpc. The combined mass of the Aurora and GS/E is in good agreement with the total mass estimated from the one-component model (\autoref{tab:summary-nogse}). For comparison, we also calculated the GS/E mass using the model of \citet{Lane2023MNRAS.526.1209L} within the same boundaries used in this paper, finding a value of $3.6\times10^7\;M_\odot$. This is approximately $1.5$ to $2.3$ times lower than the estimates from our models for the wide metallicity subsample (see \autoref{tab:summary-dpl}).

As mentioned above, the Aurora density distribution has a very steep core. However, its outskirts can be quite extended dependent on the GS/E background used. In particular, using the `DBPL+D' GS/E model for metallicities $< -2.0$, the second power index $k_2$ is so large that the Aurora mass diverges at large radii. This nuisance can probably be circumvented by applying a more sophisticated Aurora model, e.g. with a triple power-law (a double-broken one). Excessive simplicity of the model may also be a reason for the fact that the data-model residuals are not perfect for radii greater than $10$~kpc (see the Figures \ref{fig:model-two-comp-radial} and \ref{fig:model-flat-sc-1-4}, also Figures \ref{fig:model-flat-bpl} and \ref{fig:model-flat-dbpl}).

\begin{landscape}
\begin{table}
  \makegapedcells
  \setcellgapes{2pt}
  \caption{Summary results for two-component model: double power-law \eqref{eq:density-dpl} and one of the GS/E density profiles of \citet{Lane2023MNRAS.526.1209L}), see Sec. \ref{sec:dpl}. The first column is the log-likelihood related to the sample size. The last two columns are the Aurora mass and the GS/E mass estimates, correspondingly.}
  \centering
  \settowidth\rotheadsize{\theadfont Sharp core SPL}
  \begin{tabular}{c|cccccccccc}
    \hline
    Metallicity bin
    & \makecell{Log-likelihood\\$(\ln L')/N$}
    & \makecell{X-Y scale\\$a$ [kpc]}
    & \makecell{Z scale\\$b$ [kpc]}
    & \makecell{Power index\\$k_1$}
    & \makecell{Break\\$m_1$}
    & \makecell{Power index\\$k_2$}
    & \makecell{Aurora mass\\$M_\mathrm{Au}$ [$M_\odot$]}
    & \makecell{GS/E mass\\$M_\mathrm{GS/E}$ [$M_\odot$]}
    & \makecell{Total mass\\$M_\mathrm{Au} + M_\mathrm{GS/E}$ [$M_\odot$]}
    \\
    \hline
    \hline
    & \multicolumn{8}{c}{\textit{Exponentially truncated single power-law GS/E (`SC' in the paper of \cite{Lane2023MNRAS.526.1209L})}}
    \\[-4pt]
    \makecell[t{{c}}]{%
        $-3.0 \;\cdots\, -1.3$ \\[2pt]
        $-3.0 \;\cdots\, -2.0$ \\[2pt]
        $-2.0 \;\cdots\, -1.5$ \\[2pt]
        $-1.5 \;\cdots\, -1.3$ \\[2pt]
        $-1.3 \;\cdots\, -1.0$
    }
    & \makecell[t{{r}}]{$-11.477$ \\[2pt] $-11.931$ \\[2pt] $-11.443$ \\[2pt] $-11.366$ \\[2pt] $-11.164$}
    & \makecell[t{{r}}]{$2.79\pm0.01$ \\[2pt] $6.90\pm0.15$ \\[2pt] $2.91\pm0.02$ \\[2pt] $3.92\pm0.38$ \\[2pt] $3.26\pm0.02$}
    & \makecell[t{{r}}]{$1.33\pm0.01$ \\[2pt] $3.88\pm0.15$ \\[2pt] $1.38\pm0.02$ \\[2pt] $1.84\pm0.38$ \\[2pt] $1.30\pm0.02$}
    & \makecell[t{{r}}]{$4.48\pm0.01$ \\[2pt] $6.32\pm0.19$ \\[2pt] $4.51\pm0.01$ \\[2pt] $5.58\pm0.54$ \\[2pt] $5.67\pm0.01$}
    & \makecell[t{{r}}]{$\mathdash$ \\[2pt] $1.13\pm0.14$ \\[2pt] $\mathdash$ \\[2pt] $0.87\pm0.36$ \\[2pt] $2.20\pm0.04$}
    & \makecell[t{{r}}]{$\mathdash$ \\[2pt] $6.81\pm0.07$ \\[2pt] $\mathdash$ \\[2pt] $3.76\pm0.01$ \\[2pt] $4.12\pm0.01$}
    & \makecell[t{{r}}]{$2.01\times10^8$ \\[2pt] $1.09\times10^7$ \\[2pt] $9.64\times10^7$ \\[2pt] $9.04\times10^7$ \\[2pt] $2.52\times10^8$}
    & \makecell[t{{r}}]{$8.12\times10^7$ \\[2pt] $1.95\times10^7$ \\[2pt] $3.52\times10^7$ \\[2pt] $2.41\times10^7$ \\[2pt] $4.64\times10^7$}
    & \makecell[t{{r}}]{$2.83\times10^8$ \\[2pt] $3.04\times10^7$ \\[2pt] $1.32\times10^8$ \\[2pt] $1.15\times10^8$ \\[2pt] $2.98\times10^8$}
    \\
    \hline
    & \multicolumn{8}{c}{\textit{Broken power-law GS/E (`BPL')}}
    \\[-4pt]
    \makecell[t{{c}}]{%
        $-3.0 \;\cdots\, -1.3$ \\[2pt]
        $-3.0 \;\cdots\, -2.0$ \\[2pt]
        $-2.0 \;\cdots\, -1.5$ \\[2pt]
        $-1.5 \;\cdots\, -1.3$ \\[2pt]
        $-1.3 \;\cdots\, -1.0$
    }
    & \makecell[t{{r}}]{$-11.479$ \\[2pt] $-11.937$ \\[2pt] $-11.445$ \\[2pt] $-11.368$ \\[2pt] $-11.164$}
    & \makecell[t{{r}}]{$2.52\pm0.01$ \\[2pt] $7.39\pm0.18$ \\[2pt] $2.68\pm0.01$ \\[2pt] $3.90\pm0.19$ \\[2pt] $3.47\pm0.19$}
    & \makecell[t{{r}}]{$1.18\pm0.01$ \\[2pt] $4.14\pm0.18$ \\[2pt] $1.26\pm0.01$ \\[2pt] $1.76\pm0.19$ \\[2pt] $1.40\pm0.19$}
    & \makecell[t{{r}}]{$4.20\pm0.01$ \\[2pt] $6.19\pm0.24$ \\[2pt] $4.27\pm0.01$ \\[2pt] $5.87\pm0.24$ \\[2pt] $5.75\pm0.26$}
    & \makecell[t{{r}}]{$\mathdash$ \\[2pt] $1.10\pm2.03$ \\[2pt] $\mathdash$ \\[2pt] $1.07\pm0.15$ \\[2pt] $0.94\pm0.18$}
    & \makecell[t{{r}}]{$\mathdash$ \\[2pt] $3.41\pm0.04$ \\[2pt] $\mathdash$ \\[2pt] $3.56\pm0.01$ \\[2pt] $4.08\pm0.00$}
    & \makecell[t{{r}}]{$2.05\times10^8$ \\[2pt] $1.50\times10^7$ \\[2pt] $9.81\times10^7$ \\[2pt] $9.16\times10^7$ \\[2pt] $2.55\times10^8$}
    & \makecell[t{{r}}]{$7.49\times10^7$ \\[2pt] $1.31\times10^7$ \\[2pt] $3.25\times10^7$ \\[2pt] $1.76\times10^7$ \\[2pt] $5.03\times10^7$}
    & \makecell[t{{r}}]{$2.80\times10^8$ \\[2pt] $2.82\times10^7$ \\[2pt] $1.31\times10^8$ \\[2pt] $1.09\times10^8$ \\[2pt] $3.05\times10^8$}
    \\
    \hline
    & \multicolumn{8}{c}{\textit{Double-broken power-law GS/E with disc contamination (`DBPL+D')}}
    \\[-4pt]
    \makecell[t{{c}}]{%
        $-3.0 \;\cdots\, -1.3$ \\[2pt]
        $-3.0 \;\cdots\, -2.0$ \\[2pt]
        $-2.0 \;\cdots\, -1.5$ \\[2pt]
        $-1.5 \;\cdots\, -1.3$ \\[2pt]
        $-1.3 \;\cdots\, -1.0$
    }
    & \makecell[t{{r}}]{$-11.478$ \\[2pt] $-11.937$ \\[2pt] $-11.447$ \\[2pt] $-11.370$ \\[2pt] $-11.165$}
    & \makecell[t{{r}}]{$2.95\pm0.02$ \\[2pt] $8.12\pm0.12$ \\[2pt] $3.53\pm0.03$ \\[2pt] $4.25\pm0.03$ \\[2pt] $4.37\pm0.16$}
    & \makecell[t{{r}}]{$1.55\pm0.02$ \\[2pt] $5.13\pm0.13$ \\[2pt] $1.78\pm0.03$ \\[2pt] $2.02\pm0.03$ \\[2pt] $1.82\pm0.16$}
    & \makecell[t{{r}}]{$4.55\pm0.01$ \\[2pt] $7.33\pm0.17$ \\[2pt] $5.58\pm0.03$ \\[2pt] $6.89\pm0.03$ \\[2pt] $7.63\pm0.24$}
    & \makecell[t{{r}}]{$3.71\pm0.02$ \\[2pt] $1.40\pm0.13$ \\[2pt] $2.06\pm0.03$ \\[2pt] $1.98\pm0.04$ \\[2pt] $0.85\pm0.15$}
    & \makecell[t{{r}}]{$2.09\pm0.04$ \\[2pt] $0.21\pm1.19$ \\[2pt] $6.71\pm0.05$ \\[2pt] $3.14\pm0.02$ \\[2pt] $4.25\pm0.00$}
    & \makecell[t{{r}}]{$2.06\times10^8$ \\[2pt] $2.11\times10^7$ \\[2pt] $6.64\times10^7$ \\[2pt] $7.86\times10^7$ \\[2pt] $2.23\times10^8$}
    & \makecell[t{{r}}]{$5.32\times10^7$ \\[2pt] $7.11\times10^6$ \\[2pt] $5.93\times10^7$ \\[2pt] $3.25\times10^7$ \\[2pt] $5.53\times10^7$}
    & \makecell[t{{r}}]{$2.60\times10^8$ \\[2pt] $2.82\times10^7$ \\[2pt] $1.26\times10^8$ \\[2pt] $1.11\times10^8$ \\[2pt] $2.78\times10^8$}
    \\
    \hline
  \end{tabular}
  \label{tab:summary-dpl}
\end{table}
\end{landscape}

When the two-component model is applied to wide-metallicity sample, we recover essentially the same power-law slope ($k_1 = 4.48$ for the Aurora component (`SC' model, see \autoref{tab:summary-dpl}) as that obtained for the N-rich stars in \citet{Horta2021MNRAS.500.5462H} who get the power-law index of $4.47$. This agreement is in line with the hypothesis proposed in \citet{Belokurov2023} where the N-rich (or equivalently high-[N/O]) stars are predominantly members of the Aurora population. They argue that the fractional contribution of GCs to the Galactic star formation during the pre-disc, Aurora era was much larger than today and that the majority of the GCs inside the Solar radius were born in-situ. Note, however, that the model of \citet{Horta2021MNRAS.500.5462H} has a much more compact core than ours, its $z$-scale being $0.47$~kpc while the core scale in the `SC' model is almost three times larger.

We have also applied our two-component model to the sharp core Aurora profile \eqref{eq:density-spl}. This distribution converges to a single power-law away from the core. For all tested GS/E backgrounds in all metallicity bins, the profile \eqref{eq:density-spl} reached relatively low likelihoods. In some cases a degeneracy in parameters was observed: the solver tried to increase $a$, $b$ and $k_1$ indefinitely while keeping the $a/b$ and $k_1/b$ ratios constant. This can be interpreted as an attempt to extend the conical core of the profile over the entire domain volume. For these reasons we have not included results of the two-component model calculations with the profile \eqref{eq:density-spl} in the present Paper.

The Galactic coordinate frame was used to define the spatial mesh in the numerical implementation of our model. The mesh consisted of $12\,288$ pixels on the celestial sphere (using the HEALPix nested scheme of order $5$) and $41$ cells in the radial grid. This setup resulted in a cell resolution of approximately $(1.83^\circ)^2 \times 0.43\text{~kpc}$. We tested the stability of our results by varying the angular resolution of the grid. Specifically, we applied the two-component model with the `SC' GS/E to the wide metallicity subsample (similar to the model in the top row of \autoref{tab:summary-dpl}), using HEALPix schemes from orders $3$ to $6$. Our experiments showed that the spatial scales of the Aurora distribution were quite robust against changes in the grid's angular resolution. Aurora mass estimates varied by only up to $10$ per cent relative to the values in \autoref{tab:summary-dpl}. The total Aurora+GS/E mass estimate showed a variation of up to $30$ per cent. The inner slope $k_1$ became about $5$ per cent steeper at lower resolutions, making the Aurora distribution more compact. Correspondingly, the GS/E gained more mass at lower resolutions, although the total mass remained relatively unchanged.

\begin{figure}
  \centering
  \includegraphics[width=0.95\columnwidth]{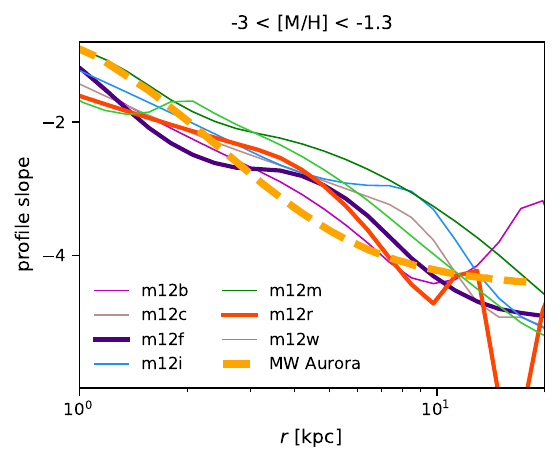}
  \caption{Comparison of the logarithmic slope of the density profile of the derived Aurora component in the DPL + SC GS/E model (orange dashed line corresponding to the orange line in the inset of Fig.~\ref{fig:model-radial-horta}) to the corresponding slopes of the in-situ pre-disk component in the MW-sized galaxies simulated as part of the FIRE-2 suite selected to have the same metallicity range of -3<[Fe/H]<-1.3 as the MW stars (solid coloured lines; the two objects {\tt m12f} and {\tt m12r} that are closest to the MW in the age-metallicity sequence \citealt{Belokurov2024gc} are shown by thicker lines).}
  \label{fig:rho_au_slope_comp}
\end{figure}

\section{Comparison with simulations}
\label{sec:simcomp}

The pre-disk Aurora component in MW-sized galaxies was predicted in cosmological simulations of galaxy formation \citep[][]{Belokurov2022MNRAS.514..689B}. Therefore, it is interesting to compare the spatial distribution of this component in simulations to the distribution derived in this study. 

Figure~\ref{fig:rho_au_slope_comp} shows a comparison of the logarithmic slope of the density profile of the Aurora component derived in this study (the DPL + SC GS/E model, shown by the thick dashed orange line)  to the corresponding slopes of the in-situ pre-disk component  in the MW-sized galaxies simulated as part of the FIRE-2 suite \citep[solid curves][]{Wetzel2023}. The in-situ stellar particles in the simulations were selected to have the same metallicity range of -3<[Fe/H]<-1.3 as the MW stars used to construct the DPL + SC GS/E model profile. 

The figure shows that simulations predict radial density profiles of the Aurora component quite close to the profile derived in the DPL+SC GS/E model. Both profiles are centrally concentrated with the slope rapidly changing from $\approx -1$ to $-1.5$ at $r=1$ kpc to $<-4$ at $r\gtrsim 8$ kpc. Although the form may seem to be close to the Hernquist profile \citep{Hernquist.1990} that describes stellar density profiles of spheroidal systems well, in the simulations the profile slope becomes shallower than -1 at $r<1$ kpc and steeper than -4 at $r>10$ kpc and thus does not correspond to this profile form. 

Note also that different simulated MW-sized galaxies have profiles remarkably close to each other despite the differences in assembly histories and halo and stellar masses. This indicates that the spatial distribution of the Aurora stars derived in this work is a generic result of galaxy formation processes. The small scatter among simulated profiles hints at a universal process that shapes the form of this profile. 

There are also small differences between the simulated profiles and the profile derived for the Aurora stars of the Milky Way. The profiles in simulations are somewhat steeper both at $r<2$ kpc and at $r>10$ kpc. Such a difference, however, can arise simply due to the specific analytic profile model adopted to derive the Aurora MW profile. For example, this profile is forced to have a slope of zero at small radii and an asymptotic slope at large radii. 

We also find that the stellar mass of the in-situ stars in the metallicity range of $\rm -3 < [Fe/H] < -1.3$ in the FIRE-2 simulations is a factor of $\approx 2-3$ larger than derived in this paper. This does not necessarily mean that simulations overproduce stellar mass or that the Aurora stellar mass is underestimated in our analysis. Given that stars are selected in a specific metallicity range, the difference in stellar mass may simply arise if simulated galaxies follow a different track in the stellar mass-metallicity plane due, for example, to the specific feedback implementation and corresponding outflows of mass and heavy elements. 
Despite this difference in normalization, we find that the density profile shape of in-situ stars in simulations does not change significantly if we vary the metallicity range used to select the stars. This makes our comparison of the {\it shape} of the profile robust.

\section{Summary and Discussion}
\label{sec:discussion}

Motivated by recent advances in the studies of the Galactic stellar halo and improvements in our understanding of the {\it Gaia} selection function, this paper presents a comprehensive model of the spatial density distribution of metal-poor giant stars in the central regions of the Galaxy. We take advantage of the stellar atmosphere properties based on the {\it Gaia} DR3 XP spectro-photometry as published recently by \citet{Andrae2023ApJS..267....8A}. We rely on their catalogue of vetted red giant stars -- this simplifies the selection process but imposes non-trivial selection effects throughout the sample. When {\it Gaia} measurements are converted into orbital properties, a clear picture emerges (see Figure~\ref{fig:data-kinem-all}) in which the Galaxy's overall spin changes dramatically around $[\mathrm{M}/\mathrm{H}] \approx -1$ in agreement with other recent studies \citep[][]{Belokurov2022MNRAS.514..689B, Rix2022, Zhang2023arXiv231109294Z,Chandra2023arXiv231013050C}. Accordingly, we focus our modelling efforts on the metallicity range $-3 < [\mathrm{M}/\mathrm{H}] < -1.3$, where the contamination from the Milky Way's disc is expected to be minimal. This is illustrated in Figure~\ref{fig:data-obs-sky} where the morphology of the projected stellar density is quasi-spheroidal at metallicities below $[\mathrm{M}/\mathrm{H}] \approx -1$ and much flatter for more metal-rich stars. We convert the number of red giant branch stars into total stellar mass using i) observed star counts in the globular cluster NGC 6397 and ii) MIST model luminosity functions and isochrones. Reassuringly, both methods yield very similar results.

We concentrate on the Galactocentric radial range $0.5<r$(kpc)$<18$ and consider two types of density models: a one-component and a two-component. For a single component, we use a vertically flattened power-law density which is modified to avoid singularity at $r=0$ and approaches the centre with a linear slope. The properties of the best-fit single-component model are displayed in Figure~\ref{fig:model-summary-nogse}. A small spatial scale of order of 1 kpc is preferred for $[\mathrm{M}/\mathrm{H}] > -2$ but increases by a factor of $\approx 3$ for lower metallicities. The density is flattened vertically with $b/a \approx 0.5$, in good agreement with a variety of studies \citep[see][and references therein]{Deason2011, Horta2021MNRAS.500.5462H}.  Figure~\ref{fig:model-summary-nogse} also shows a mildly increasing flattening with increasing metallicity, possibly signalling small amount of disc contamination at $[\mathrm{M}/\mathrm{H}] > -1.3$. Our best-fit power-law index for $[\mathrm{M}/\mathrm{H}]< -1.3$ is $-3.4$ which is consistent with other recent studies focused on the inner halo \citep[see e.g.][and references therein]{Deason2018pileup, Han2022, Lane2023MNRAS.526.1209L}, but is somewhat steeper than the stellar halo density measurements outside of $r \approx 5$ kpc \citep[see e.g.][]{Deason2011, Iorio2018}. Our one-component model gives the total stellar halo mass of $\approx 3\times10^8 M_\odot$ for $-3 < [\mathrm{M}/\mathrm{H}] < -1.3$. This estimate increases by a factor of $\approx2$ if stars with metallicities up to $[\mathrm{M}/\mathrm{H}] = -1$ are included.

We have two main incentives to consider a more complex model. First, we strive to improve the residuals of the one-component model which show some systematic trends in the very inner and outer parts of the domain considered (see Figures~\ref{fig:model-cuts-sharp-nogse-0}, \ref{fig:model-proj-sharp-nogse-0}, \ref{fig:model-radial-sharp-nogse-0}). Second, we are guided by the recent analysis of the inner Galactic halo which shows evidence for at least two individual components with distinct kinematic, chemical and spatial distributions (Davies et al 2024, in prep). In terms of the exact origin of these populations, one is attributed to the so-called {\it Gaia} Sausage/Enceladus, the last significant merger experienced by the MW around 10 Gyr ago \citep[][]{Belokurov2018MNRAS.478..611B, Helmi2018}. The precise nature of the second component remains unconstrained but it appears to be connected to the ancient, prehistoric and, likely, pre-disc state of the Milky Way. Different scenarios have been discussed in the literature, invoking an early accretion event \citep[see e.g.][]{Horta2021heracles}, rapid in-situ formation \citep[][]{Belokurov2022MNRAS.514..689B, Conroy2022, Rix2022}, possibly with significant contribution from disrupting Globular Clusters \citep[see e.g.][]{Belokurov2023}.

In the two-component analysis, we fix the GS/E contribution to one of the models reported in \citet{Lane2023MNRAS.526.1209L}. For the second component, we fit the giant count with a vertically flattened double power-law density models. Again, these models are modified to avoid singularity near origin, smoothly evolving into a flat core inside the scale radius. We proceed by trying different GS/E parametric shapes with the parameters fixed to the best-fit values in \citet{Lane2023MNRAS.526.1209L} but leaving the component's density normalization free. Unsurprisingly, once the GS/E's contribution is set, the remaining density component is revealed to have a rather steep radial density fall-off. As Table~\ref{tab:summary-dpl} shows, our best-fit models for the proto-Galaxy prefer a power-law index of $\approx-4.5$ when stars from a wide range of metallicities ($-3<$[M/H]$<-1.3$) are considered, but below [M/H]$<-1.5$ even steeper fall-offs are reported. An example of a well-behaved two-component fit is given in Figures~\ref{fig:model-two-comp-params} and \ref{fig:model-two-comp-radial}. As the latter Figure demonstrates, compared to the single-component fit, the residuals in the inner and the outer parts of the dataset are indeed improved. Figure~\ref{fig:model-two-comp-radial} makes it rather clear: outside of the Solar radius, the GS/E debris dominate the stellar halo, but the inner parts are the realm of Aurora. Around the location of the Sun, the two halo populations appear to contribute approximately equally.

The reliability of our stellar halo component decomposition can be independently verified through chemistry. Most recently, Davies et al (2024) carried out a blind source separation of the nearby stellar halo using elemental abundances measured by the APOGEE spectroscopic survey (REF). In their experiment, they assumed that in each small region of the integrals-of-motion space, more precisely, the space spanned by the stars' orbital energy and angular momentum, the distribution function is a linear combination of two distinct components. The components' behaviour is analysed in the space of abundances ratios [Al/Fe], [Mg/Fe] and [Fe/H]. Davies et al (2024) demonstrate that with Non-negative Matrix Factorisation, two components with markedly different chemical trends can be teased out of APOGEE stellar mixtures automatically. The fractional contributions of the components are strong functions of orbital energy, and invert around the Solar value. Figure~\ref{fig:model-radial-aurora-flat-sc} compares the fractional contribution of the low-energy component obtained by Davies et al (2024) and the fraction of the density component we associate with the Aurora population here. Reassuringly, in both NMF analysis of the APOGEE data and in our 3D density modelling, the low-energy/Aurora's contribution is dominant inside the Solar radius, dropping to $\approx50\%$ around the Sun's position. Better still, the two different estimates of the change of the Aurora fraction with radius agree as well.

To place our study into context, Figure~\ref{fig:model-radial-horta} compares various radial density profiles obtained here with other works in the literature. It shows that our single-component model (black dotted line) is only slightly steeper than the halo model of \citet{Horta2021MNRAS.500.5462H} shown as a solid blue line. Note however, that \citet{Horta2021MNRAS.500.5462H} include more metal-rich APOGEE giants in their halo sample. As Figure~\ref{fig:model-summary-nogse} shows, if we extended the metallicity range to [M/H]$=-1$, the resulting single-component profile would steepen slightly, likely improving the agreement further. As mentioned earlier, around the Sun, our Aurora  density profile (solid orange line) matches the power-law density of N-rich stars in \citet{Horta2021MNRAS.500.5462H}. Around $r\approx3$ kpc, our Aurora model starts to flatten to become a core close to the Galactic centre, while the reported N-rich power-law model continues to eventually diverge at the origin. We also note that outside of the Solar radius, our combined two-component model matches rather well the stellar halo density as measured by \citet{Deason2011} using A-coloured stars (Blue Horizontal Branch stars and Blue Stragglers). The agreement between the two models obtained using completely different datasets, selection procedures and tracer populations is encouraging.

\section{Data availability}

All data used in this paper is publicly available.

\section{Acknowledgements}

This work is a result of the GaiaUnlimited (\url{https://gaia-unlimited.org/}) project, which has received funding from the European Union’s Horizon 2020 research and innovation program under grant agreement No 101004110. The GaiaUnlimited project was started at the 2019 Santa Barbara Gaia Sprint, hosted by the Kavli Institute for Theoretical Physics at the University of California, Santa Barbara.

This work has made use of data from the European Space Agency (ESA) mission \textit{Gaia} (\url{https://www.cosmos.esa.int/gaia}), processed by the \textit{Gaia} Data Processing and Analysis Consortium (DPAC, \url{https://www.cosmos.esa.int/web/gaia/dpac/consortium}). Funding for the DPAC has been provided by national institutions, in particular the institutions participating in the \textit{Gaia} Multilateral Agreement.

This research or product makes use of public auxiliary data provided by ESA/Gaia/DPAC/CU5 and prepared by Carine Babusiaux.

We used FIRE-2 simulation public data  \citep[][]{Wetzel2023}, which are part of the Feedback In Realistic Environments (FIRE) project, generated using the Gizmo code \citep{hopkins15} and the FIRE-2 physics model \citep{hopkins_etal18}.


We thank Anke Ardern-Arentsen, Eugene Vasiliev, Hanyuan Zhang, David W. Hogg, and Adrian Price-Whelan for advice and valuable discussions.

\bibliography{biblio}
\bibliographystyle{mnras_alt}

\appendix

\section{Transformation of probability}
\label{sec:transformation}

\begin{figure}
  \centering
  \includegraphics[width=0.8\columnwidth]{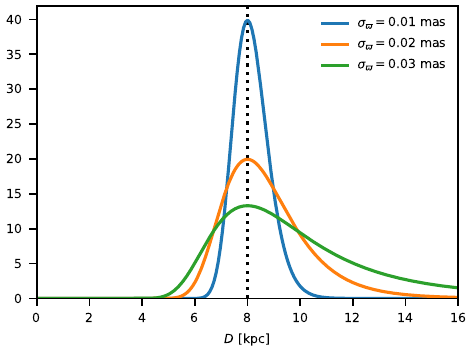}
  \caption{Distribution of the observed parallax values \eqref{eq:plx-conv-ker} shown in the distance values, $\mathcal{G}(D^{-1}\,|\,D'^{-1}, \sigma_\varpi)$, given $D' = 8$ kpc.}
  \label{fig:model-plx-conv-ker}
\end{figure}

The physical model of the stellar distribution $\rho(\vec{r} |\,\theta)$ is defined in the Galactocentric coordinate frame, while the observable positions of the stars are given in the Galactic frame. Due to parallax measurement errors, the transformation between the true coordinates of a star and the measured coordinates is stochastic. In the statistical model we implement (see Section \ref{sec:statistical-model}), it is assumed that the density distribution is described by a PDF in the positional space. This view allows the parallax errors to be included in the probability transformation function during the conversion from physical to observed density distributions. Let us look at this in more detail.

Suppose $(l', b', D')$ are the \textit{true} Galactic coordinates, and $(l, b, D)$ are the \textit{observed} coordinates. One can neglect the errors in the celectial positions of a star, so hereinafter we will assume $l = l'$ and $b = b'$. Let $\varpi' = D'^{-1}$ be the true parallax, i.e. related to the true distance to the star. Suppose the observed parallax $\varpi$ has an error whose standard deviation is $\sigma_\varpi$. The value of $\varpi$ can be thought of as a random quantity with the Gaussian distribution
\begin{equation}
  \mathcal{G}(\varpi\,|\,\varpi', \sigma_\varpi)
  = \frac{1}{\sqrt{2\pi}\,\sigma_\varpi}\,\exp\!\left[ - \frac{(\varpi - \varpi')^2}{2 \sigma_\varpi^2} \right]  \;.
  \label{eq:plx-conv-ker}
\end{equation}
The typical size of the error of \textit{Gaia} parallaxes is of order $0.01$ \citep{GaiaPerformance}. This value is crucial for distance estimation, especially since the errors that are symmetric in parallax become asymmetric in the distance, see Figure \ref{fig:model-plx-conv-ker}.

The parallax uncertainties induce the uncertainties in the density distribution. Consider $\lambda(l, b, \varpi)$ [mas$^{-1}$ sr$^{-1}$] is a linear number density of stars at the celestial position $(l, b)$ and the parallax $\varpi$, i.e. $\lambda(l, b, \varpi)\,d\varpi$ is a number of stars in the interval of parallaxes $(\varpi, \varpi + d\varpi)$ along a cone of sight with the unit solid angle around the direction $(l, b)$. The observed linear density at the position $(l, b, \varpi)$ is as follows:
\begin{equation}
  \lambda_\mathrm{obs}(l, b, \varpi)
  = \int_{-\infty}^{+\infty} d\varpi'\,\lambda(l, b, \varpi')\,\mathcal{G}(\varpi\,|\,\varpi', \sigma_\varpi)  \;.
  \label{eq:linear-plx-density}
\end{equation}
The parallax error $\sigma_\varpi$ generally depends on the celestial position and on the apparent magnitude. We will estimate this dependence in next Appendix subsection.

Let us multiply the observed linear density $\lambda_\mathrm{obs}$ by a solid angle $d\Omega$, then integrate over a parallax interval $(\varpi_1, \varpi_2)$. Obviously, we will get a number of stars inside the conical layer of the solid angle $d\Omega$:
\begin{equation}
  \int_{D_1}^{D_2} dN_\mathrm{obs}(l, b, D)
  = d\Omega \int_{\varpi_2}^{\varpi_1} d\varpi\,\lambda_\mathrm{obs}(l, b, \varpi)  \;,
\end{equation}
where $D_1$ and $D_2$ are the distances defined as the inverse parallaxes $\varpi_1$ and $\varpi_2$, respectively; $dN_\mathrm{obs}(l, b, D)$ is a number of stars observed inside a thin conical layer of a thickness $dD$ and the solid angle $d\Omega$. Using these definitions, we can rewrite the Eq.~\eqref{eq:linear-plx-density} as follows:
\begin{equation}
  dN_\mathrm{obs}(l, b, D)
  = \bigl|d(D^{-1})\bigr| \int dN(l, b, D')\,\mathcal{G}(D^{-1}\,|\,{D'}^{-1}, \sigma_\varpi)  \;,
  \label{eq:num-stars-transform}
\end{equation}
for $dN(D')$ is a \textit{true} number of stars in the conical layer enclosed by the thin distance interval $dD'$ at $D'$ and the solid angle $d\Omega$; the integration is performed over the distances $D'$.

The statistical model defined in the Section \ref{sec:statistical-model} operates with the density distribution as a probability distribution function in 3D position space. Denoting $\rho_\mathrm{obs}$ an observed PDF, one can write
\begin{equation}
  dN_\mathrm{obs}(l, b, D)
  = N_\mathrm{tot}\,\rho_\mathrm{obs}(l, b, D)\,D^2\,dD\,d\Omega  \;,
  \label{eq:num-stars-vs-pdf}
\end{equation}
where $N_\mathrm{tot}$ is a total number of stars. The similar relation takes place between the theoretical PDF $\rho(l, b, D')$ and the true number of stars $dN(l, b, D')$.

The relations \eqref{eq:num-stars-transform} and \eqref{eq:num-stars-vs-pdf} are not sufficient to write out a transform between the theoretical and the observable densities. The reason for that is the parallax error $\sigma_\varpi$ depends not only on the celestial position but also on the apparent magnitude. This can be accounted for by introducing a luminosity function into the model. Let us denote $\Phi(\mathrm{M}_\mathrm{G})$ a PDF for absolute magnitude $\mathrm{M}_\mathrm{G}$, and $f$ a PDF in space of the observable variables $(l, b, D, \mathrm{G})$, i.e. a joint PDF for a star to be observed at the position $(l, b, D)$ with an apparent magnitude $\mathrm{G}$. Considering \eqref{eq:num-stars-transform} and \eqref{eq:num-stars-vs-pdf}, we may state that
\begin{multline}
  f(l, b, D, \mathrm{G})
  = \int d\mathrm{M}_\mathrm{G}\,\Phi(\mathrm{M}_\mathrm{G}) \int dD'\,D'^2 \rho(l, b, D')  \\
    \times T(D, \mathrm{G}\,|\,l, b, D', \mathrm{M}_\mathrm{G})  \;,
  \label{eq:lbDG-joint-pdf}
\end{multline}
where $T$ is the transformation probability,
\begin{multline}
  T(D, \mathrm{G}\,|\,l, b, D', \mathrm{M}_\mathrm{G})
  = \left| \diff{(D^{-1})}{D} \right|
    \,\mathcal{G}\bigl(D^{-1}\,|\,{D'}^{-1}, \sigma_\varpi(l, b, \mathrm{G})\bigr)  \\
    \times \delta\bigl[\mathrm{M}_\mathrm{G} + A_\mathrm{G}(l, b) + 5\log_{10} D' + 10 - \mathrm{G}\bigr]  \;.
\end{multline}
Here $\delta[\dots]$ is the Dirac's delta function, and $A_\mathrm{G}(l, b)$ is the extinction in $\mathrm{G}$ band (see the Section~\ref{sec:sample-selection}). Note that the delta function depends on the distance $D'$, not $D$. This is because the apparent magnitude depends on the true parallax, not the observed one.

In the reasoning above, we have not taken into account the selection efficiency. This means that some of the stars are lost from the true distribution. The selection effects can be described with the selection function, see the Section~\ref{sec:statistical-model} and Appendix~\ref{sec:xp-selection}.

\section{Model for parallax errors}
\label{sec:parallax-error}

\setcounter{figure}{7}
\begin{figure*}
  \centering
  \includegraphics[width=\textwidth]{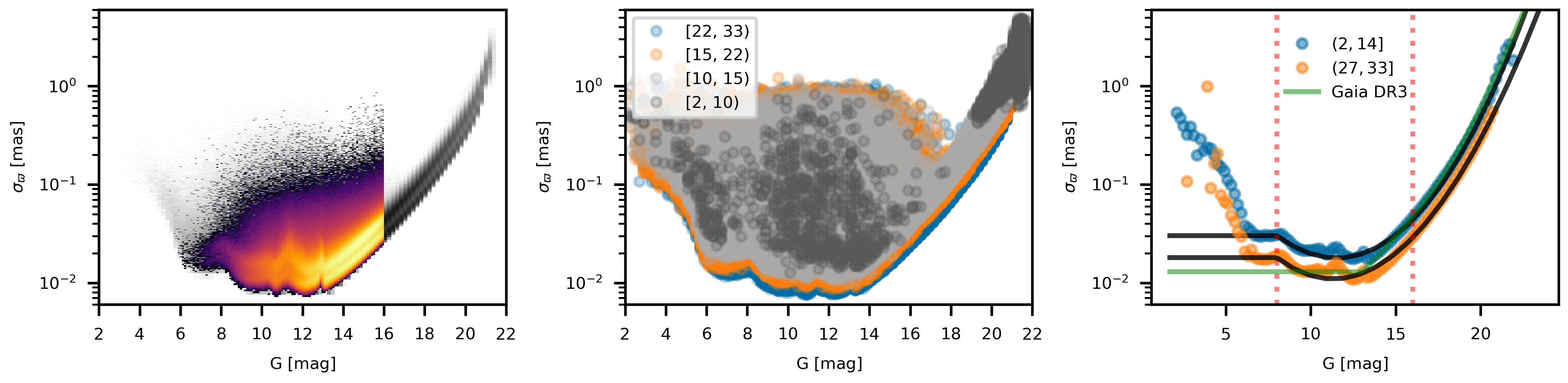}
  \caption{\textit{Left panel} are parallax errors vs. apparent magnitudes of RGB stars (red-and-yellow colour map) and all the \texttt{gaia\_source} stars (grey colour map on the background). \textit{Middle panel} are the parallax errors vs. apparent magnitudes of all the \texttt{gaia\_source} stars binned in \texttt{visibility\_periods\_used} (see the legend). \textit{Right panel} are the parallax error vs. apparent magnitude in two bins of \texttt{visibility\_periods\_used} (see a legend on the plot) estimated for all the \texttt{gaia\_source} stars. Black lines are the fit. The green curve is the Gaia DR3 astrometric performance model \citep{GaiaPerformance}. A left dotted line at $\mathrm{G} = 8$ shows the saturation point of the $\sigma_\varpi \mathdash \mathrm{G}$ dependency for RGB stars. A right dotted line at $\mathrm{G} = 16$ is the limiting magnitude for RGB stars in the \citet{Andrae2023ApJS..267....8A} catalogue.}
  \label{fig:plxerr}
\end{figure*}

\begin{figure}
  \centering
  \includegraphics[width=\columnwidth]{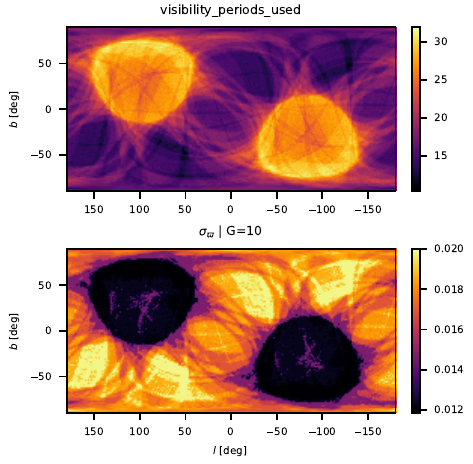}
  \caption{\textit{Top panel}: Distribution of \texttt{visibility\_periods\_used} over the sky (a median values per HEALpix of level 7). \textit{Bottom panel}: Model for parallax errors for $\mathrm{G} = 10$.}
  \label{fig:plxerr-sky}
\end{figure}

The density smearing noted in the Sec.\ref{sec:statistical-model} and Appendix~\ref{sec:transformation} is determined by the errors in measuring of parallaxes. The parallax error depends on instrumental and calibration effects, and has a trend strongly depending on the apparent magnitude \citep{GaiaPerformance} (see the left panel at the Figure \ref{fig:plxerr}). It is also reasonable to expect that the parallax error will depend on the number of observations of a particular star: errors should be smaller if the observations are as far apart in time. To test these considerations, we bin the parallax errors with the \texttt{visibility\_periods\_used} values from the \texttt{gaia\_source} table. As seen on the Figure \ref{fig:plxerr} (middle panel), variations in the number of visibility periods for the same apparent magnitude may lead to the variations in the parallax error almost to the factor three (note the log scale on the figure). It may substantially affect the parallax smearing (Figure \ref{fig:model-plx-conv-ker}).

In order to evaluate the $\sigma_\varpi(l, b, \mathrm{G})$ dependence, we estimated the median \texttt{parallax\_error} and median \texttt{visibility\_periods\_used} in every HEALPix 7 on the sky and also in $\mathrm{G}$ bins. Next, the parallax errors were grouped by the \texttt{visibility\_periods\_used} values (the last was binned at the edges $2$, $14$, $15$, $16$, $17$, $19$, $20$, $24$, $26$, $27$, and $33$) then fitted as a function of $\mathrm{G}$. It is important to note that in the sample of RGB stars we used, the parallax error looks saturated for the stars brighter than $\mathrm{G} = 8$ mag, comparing with the whole \texttt{gaia\_source} catalogue (grey shaded area on the Figure \ref{fig:plxerr}). While such stars are unlikely in the sample (Figure \ref{fig:data-magnitudes}), these magnitudes are possible in the model. In the fit, the saturation of the parallax error was forced for $\mathrm{G} \leq 8$. Resulting fits are shown on the Figure \ref{fig:plxerr} (right panel), together with the \cite{GaiaPerformance} astrometric performance model, which is sky averaged.

Using the median number of the visibility periods estimate for every HEALPix, it is possible to make a map of the parallax error as a function of a sky position and of the apparent magnitude. An example for $\mathrm{G} = 10$ is shown in a Figure \ref{fig:plxerr-sky}. Note the error rising in the low visibility areas.

\section{XP stars selection function}
\label{sec:xp-selection}

\begin{figure}
  \centering
  \includegraphics[width=\columnwidth]{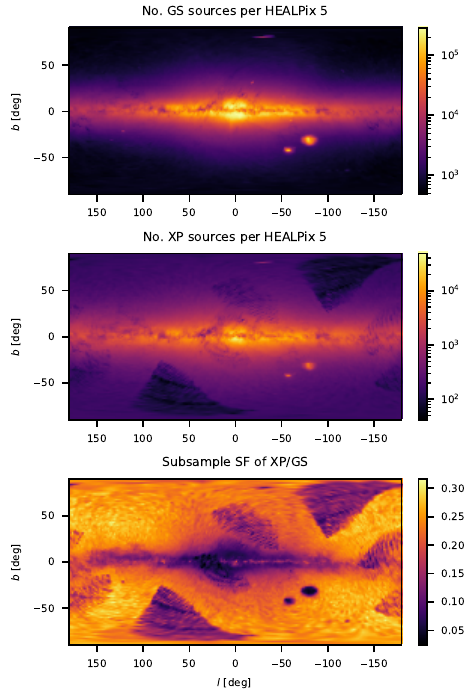}
  \caption{\textit{From top to bottom}: count of stars of the whole \texttt{gaia\_source} (GS) catalogue inside HEALPixels of the level 5, count of stars which in the \citet{Andrae2023ApJS..267....8A} catalogue (RGB stars having XP spectra), the second panel divided by the first panel.}
  \label{fig:data-sfxp-sky}
\end{figure}

\begin{figure}
  \centering
  \includegraphics[width=0.42\textwidth]{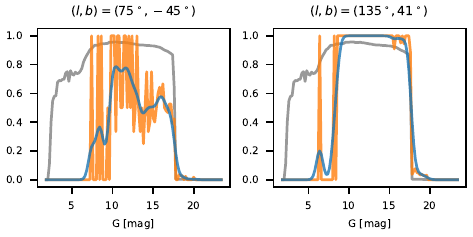}
  \caption{Subsample selection function per apparent magnitude bin (of $0.2$ mag wide), for along the two sky directions. \textit{Grey line} is the overall sky MLE. \textit{Orange} is the MLE along a specified direction $(l, b)$. \textit{Blue} is the same after the Gaussian smoothing (using $0.4$ mag standard deviation).}
  \label{fig:data-sfxp}
\end{figure}

The observed distribution of stars on the sky in the Figure \ref{fig:data-obs-sky} reveals notable gaps not only in the Galactic plane but also in regions of low visibility ($50^\circ < l < 100^\circ$, $b < -20^\circ$ and in the opposite direction). The last is due to the scanning law of the \textit{Gaia}, whose peculiarities led to a relative lack of observations in these areas.

The catalogue of RGB stars of \cite{Andrae2023ApJS..267....8A} we rely on is subsampled from the catalogue of the stars that have XP spectra. A simple way to estimate this selection effect is to relate a number of stars with XP spectra to a number of all stars in the \texttt{gaia\_source} catalogue, see Figure \ref{fig:data-sfxp-sky}. In this way, we can obtain a Maximum Likelihood Estimate (MLE) for the selection probability in the HEALPixels and the $\mathrm{G}$ bins.

The catalogue of RGB stars of \cite{Andrae2023ApJS..267....8A} we rely on is subsampled from the catalogue of the stars that have XP spectra. Adhering to the probabilistic approach we used in Section~\ref{sec:statistical-model}, the efficiency of the subsampling or the probability of selection can be defined as a probability for a successful Bernoulli trial \citep{Boubert2020MNRAS.497.4246B, Castro-Ginard2023A&A...677A..37C}. A simple way to estimate the probability is to relate a number of stars with XP spectra to a number of all stars in the \texttt{gaia\_source} catalogue%
\footnote{%
An alternative approach is to evaluate the empirical selection function by a method of \citet{Cantat-Gaudin2023A&A...669A..55C}.
}.%
In this way, we can obtain a Maximum Likelihood Estimate (MLE) for the selection probability in the HEALPixels and the $\mathrm{G}$ bins. Figure~\ref{fig:data-sfxp-sky} shows the sky projection of the source counts in both the parent and XP catalogue, and also their ratio.

When estimating the selection function in the apparent magnitude range, the dependence on $\mathrm{G}$ seems to be essentially noisy. This may be due to the randomness of the distribution of stars which is particularly evident in subsamples within small solid angle. Two examples of this are shown on a Figure \ref{fig:data-sfxp}: a low visibility pixel, and a high visibility one.

To overcome the unwanted statistical noise, the selection function was smoothed in $\mathrm{G}$ with the Gaussian kernel of the width $0.4$ mag.

\section{Additional figures}
\label{sec:add-figs}

\begin{figure*}
  \centering
  \includegraphics[width=0.42\textwidth]{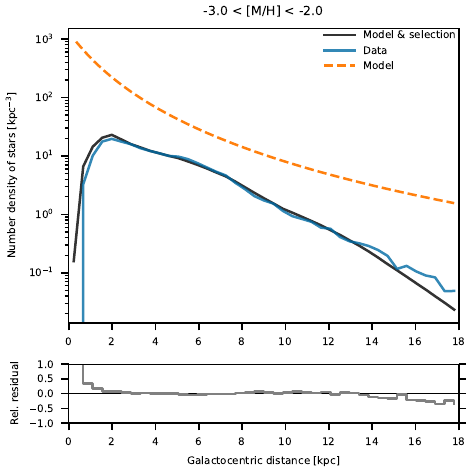}
  \includegraphics[width=0.42\textwidth]{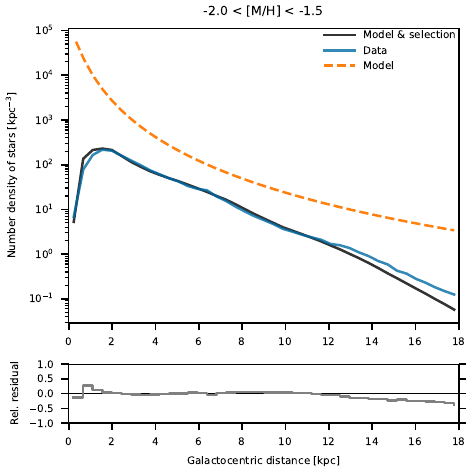}
  \includegraphics[width=0.42\textwidth]{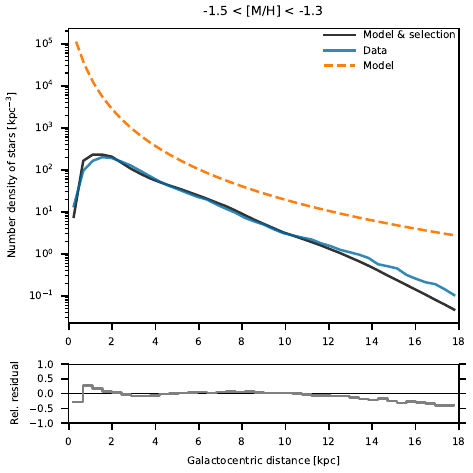}
  \includegraphics[width=0.42\textwidth]{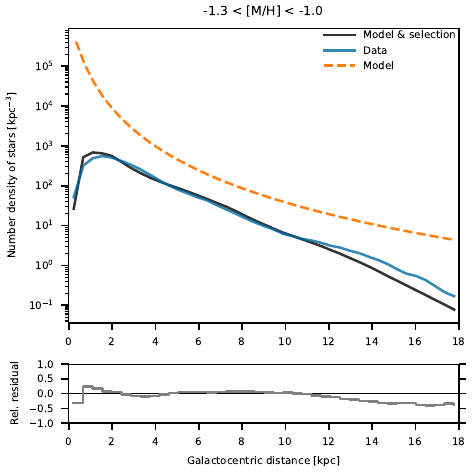}
  \caption{Single-component model. Radial profiles of the number density of stars for a set of metallicity bins in the single component no-GS/E model. Designations for lines are the same as in Figure \ref{fig:model-two-comp-radial}.}
  \label{fig:model-radial-sharp-nogse-1-4}
\end{figure*}

\begin{figure*}
  \centering
  \includegraphics[width=0.42\textwidth]{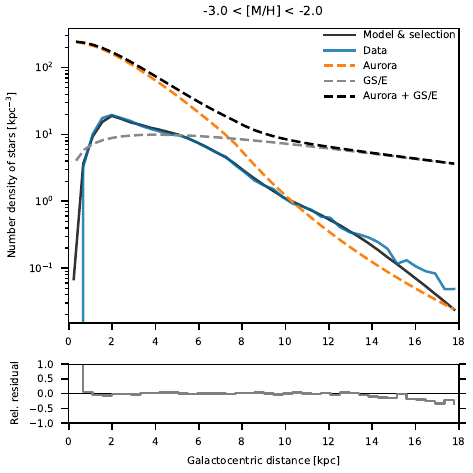}
  \includegraphics[width=0.42\textwidth]{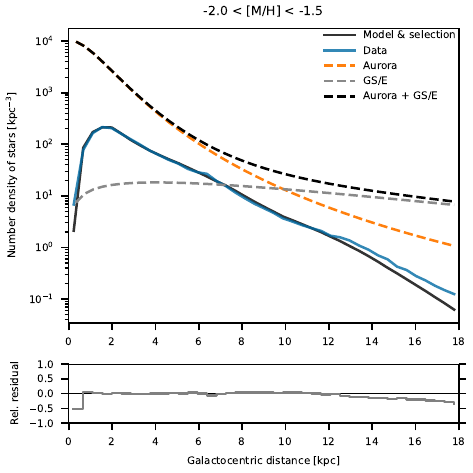}
  \includegraphics[width=0.42\textwidth]{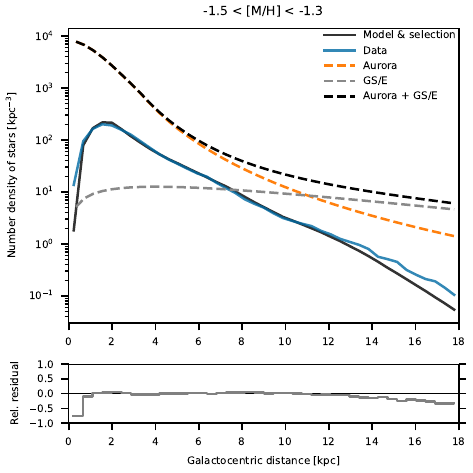}
  \includegraphics[width=0.42\textwidth]{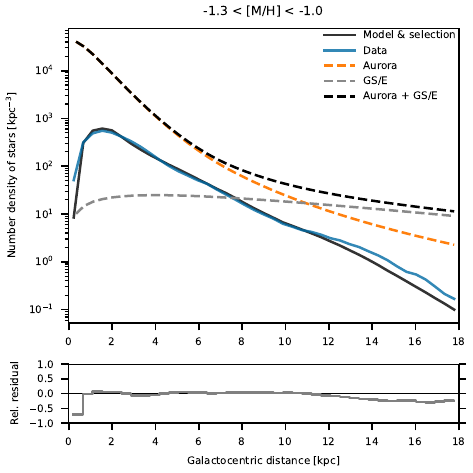}
  \caption{DPL Aurora \& SC GS/E. Radial density profiles for a set of metallicity bins.  Designations are the same as in Figure \ref{fig:model-two-comp-radial}.}
  \label{fig:model-flat-sc-1-4}
\end{figure*}

\begin{figure*}
  \centering
  \includegraphics[width=0.42\textwidth]{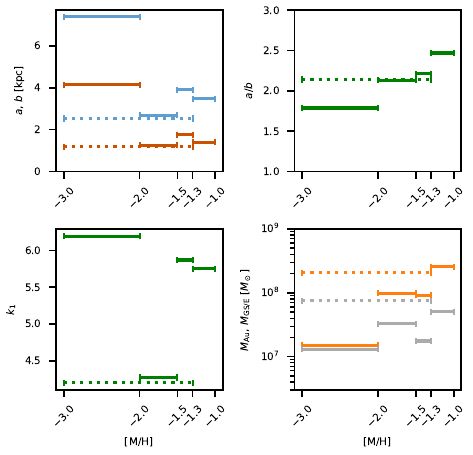}
  \includegraphics[width=0.42\textwidth]{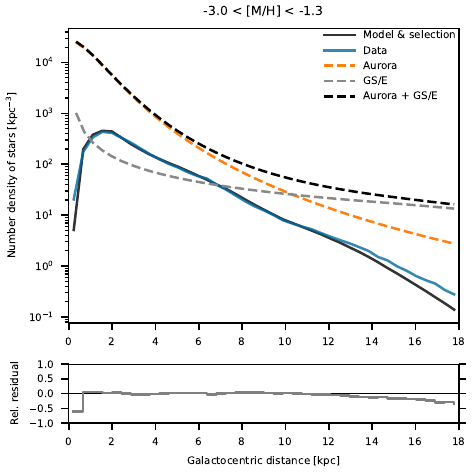}
  \\
  \includegraphics[width=0.42\textwidth]{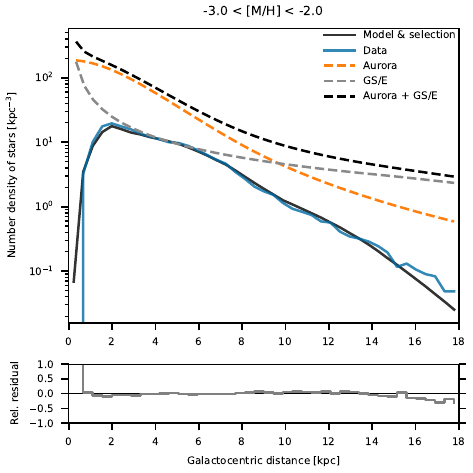}
  \includegraphics[width=0.42\textwidth]{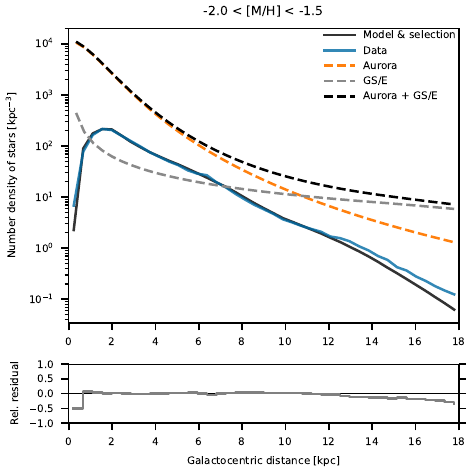}
  \includegraphics[width=0.42\textwidth]{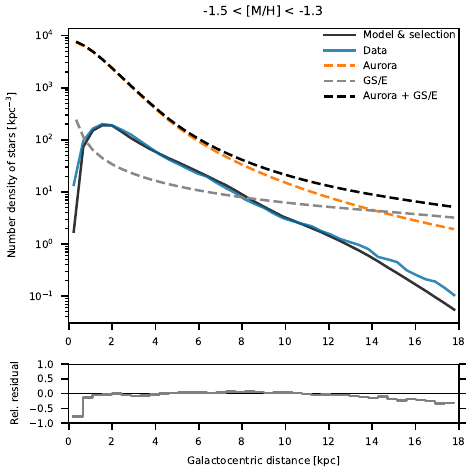}
  \includegraphics[width=0.42\textwidth]{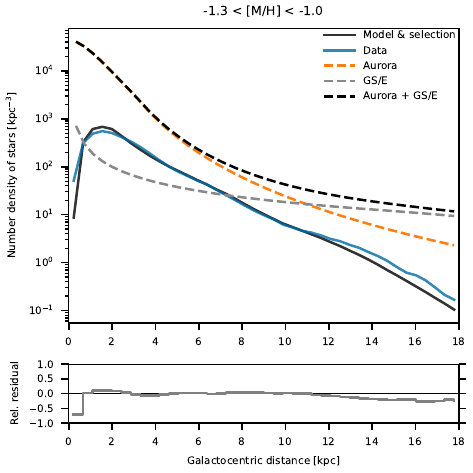}
  \caption{DPL Aurora \& BPL GS/E. All line designations are the same as in Figures \ref{fig:model-two-comp-params} and \ref{fig:model-two-comp-radial}.}
  \label{fig:model-flat-bpl}
\end{figure*}

\begin{figure*}
  \centering
  \includegraphics[width=0.42\textwidth]{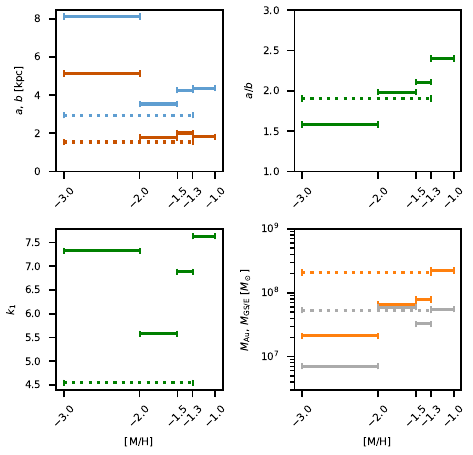}
  \includegraphics[width=0.42\textwidth]{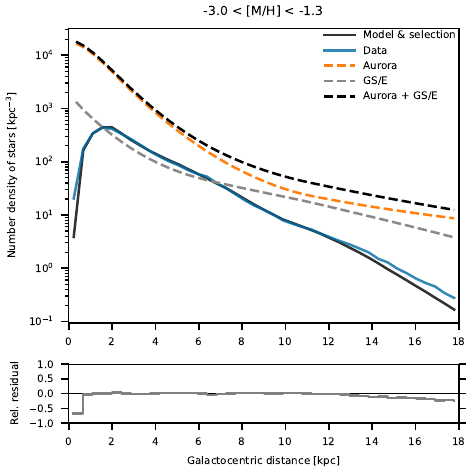}
  \\
  \includegraphics[width=0.42\textwidth]{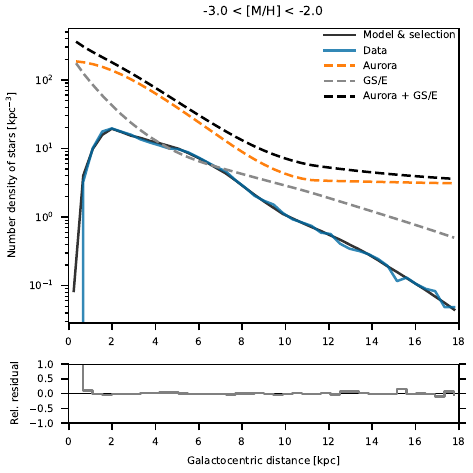}
  \includegraphics[width=0.42\textwidth]{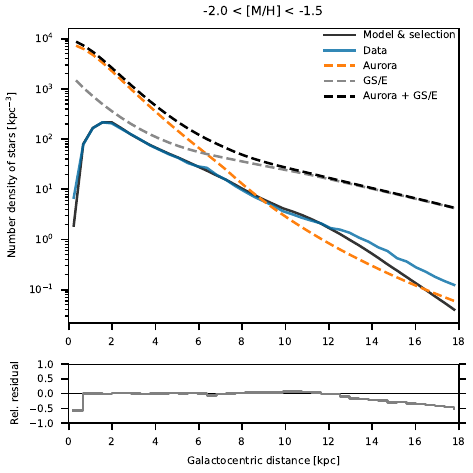}
  \includegraphics[width=0.42\textwidth]{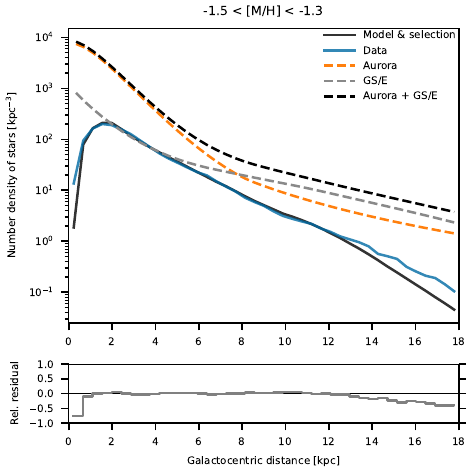}
  \includegraphics[width=0.42\textwidth]{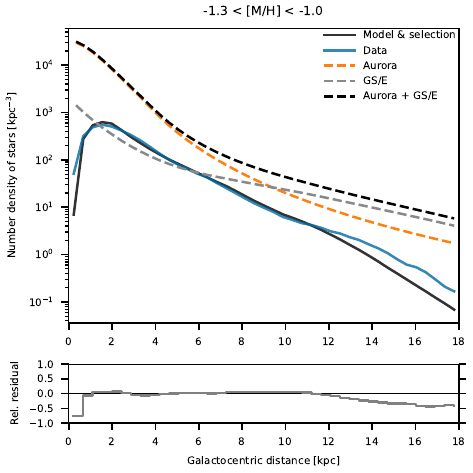}
  \caption{DPL Aurora \& DBPL+D GS/E. All line designations are the same as in Figures \ref{fig:model-two-comp-params} and \ref{fig:model-two-comp-radial}.}
  \label{fig:model-flat-dbpl}
\end{figure*}

The Figure~\ref{fig:model-radial-sharp-nogse-1-4} are radial profiles for the one-component model (Sec.~\ref{sec:spl-nogse}). The Figures~\ref{fig:model-flat-sc-1-4}, \ref{fig:model-flat-bpl} and \ref{fig:model-flat-dbpl} are of the two-component models (Sec.~\ref{sec:dpl})

\end{document}